\documentclass[12pt,fleqn]{article}
\usepackage{psfig,float}
\begin{document}
\begin{center}
{\Large{\bf The role of $\Delta$(1700) excitation and $\rho$ }}
\end{center}
\begin{center}
{\Large{\bf production in double pion}}
\end{center}
\begin{center}
{\Large{\bf photoproduction}}
\end{center}
\vspace{1cm}

\begin{center}
{\large{ J.C. Nacher, E. Oset, M.J. Vicente Vacas and L. Roca}}
\end{center}
\vspace{0.4cm}
\begin{center}
{\it Departamento de F\'{\i}sica Te\'orica and IFIC}
\end{center}
\begin{center} 
{\it Centro Mixto Universidad de Valencia-CSIC}
\end{center}
\begin{center}
{\it Institutos de Investigaci\'on de Paterna, Apdo. correos 22085,}
\end{center}
\begin{center}
{\it 46071, Valencia, Spain}       
\end{center}

\vspace{3cm}

\begin{abstract}
{\small{Recent information on invariant mass distributions of the 
$\gamma p\rightarrow\pi^+\pi^0 n$ reaction, where previous theoretical
models had shown deficiencies, have made more evident the need for new
mechanisms, so far neglected or inaccurately included. We have updated a
previous model to include new necessary mechanisms. We find that the production
of the $\rho$ meson, and the $\Delta(1700)$ excitation, through interference
with the dominant terms, are important mechanisms that solve the puzzle of the
$\gamma p\rightarrow\pi^+\pi^0 n$ reaction without spoiling the early agreement
with the $\gamma p\rightarrow\pi^+\pi^- p$ and 
$\gamma p\rightarrow\pi^0\pi^0 p$ reactions.}}
\end{abstract}
\newpage

\section{Introduction}
 Recently, new improvements in the experimental techniques have made possible
the study of total cross sections with accuracy 
for the two pion photoproduction reactions as:
$\gamma p\rightarrow\pi^+\pi^- p$, $\gamma p\rightarrow\pi^0\pi^0 p$,
$\gamma p\rightarrow\pi^+\pi^0 n$ and $\gamma n\rightarrow\pi^-\pi^0 p$
using the large acceptance detector DAPHNE and high intensity
tagged photons at Mainz. Some polarization observables are being also measured, 
 like the  spin assymetry $\sigma_{3/2}
-\sigma_{1/2}$ and the helicity cross sections $\sigma_{1/2}$ and
$\sigma_{3/2}$ with the DAPHNE acceptance \cite{lang}. The invariant masses  of $\pi^0\pi^0$
\cite{wolf}, $\pi^-\pi^0$ \cite{za2}
and $\pi^+\pi^0$ \cite{metag} have also been measured for different bins of incident photon 
energies.

This new wealth of data has stimulated us to search for missing mechanisms in
previous theoretical models in order to find a suitable description of the
different observables in all those channels.

 A model for the $\gamma p\rightarrow\pi^+\pi^- p$ was developed in
\cite{tejedor} finding a good reproduction of the cross section up to about
$E_\gamma = 1$ GeV. A more reduced set of Feynman diagrams was found 
sufficient to
describe the reaction up to $E_\gamma \simeq 800$ MeV \cite{tejedor2} 
where the Mainz
experiments are done \cite{bra,za,ha}. In the work \cite{tejedor2} the model is 
extended
to all six isospin channels $\gamma N\rightarrow\pi\pi N$, 
and provides a quite good reproduction of the experimental
results for the $\gamma p\rightarrow\pi^+\pi^- p$ and 
$\gamma p\rightarrow\pi^0\pi^0 p$ channels. However, that model underestimates
the total cross section of the other two measured double photoproduction
channels $\gamma n\rightarrow\pi^-\pi^0 p$ and 
$\gamma p\rightarrow\pi^+\pi^0 n$, the last one by about 40 $\%$.

Other models have been proposed. In \cite{laget} a model which contains the dominant
terms of \cite{tejedor} plus some extra terms, which only become relevant at high
energies, like the $\Delta(1700)$ excitation, is shown. The model obtains a reasonable description for the
$\gamma p\rightarrow\pi^+\pi^- p$ channel but fails in the $\pi^0\pi^0$ channel 
and in
the channels in which \cite{tejedor2} is failing too. 
A revision of this work is under consideration \cite{laget2}. The model of \cite{ochi} has less diagrams than the one of
 \cite{tejedor,tejedor2} but introduces the 
$N^\ast(1520)\rightarrow N \rho$ decay mode. By fitting a few parameters to 
$(\gamma, \pi \pi)$ data the cross sections are reproduced,
including the $\gamma p\rightarrow\pi^+\pi^0 n$ and
$\gamma n\rightarrow\pi^-\pi^0 p$ reactions where the models 
of \cite{tejedor2,laget} fail.
 
The model of \cite{ochi} fails to reproduce some invariant mass distributions
 where the model of
\cite{tejedor} shows no problems, but a different version of the model of
\cite{ochi} is given 
 in \cite{ochi2,ochi3} where
the parameters of the model are changed in order to reproduce also 
the mass distribution, 
without  spoiling the cross sections. 
The $(\gamma,\pi^0 \pi^0)$ channel is somewhat underpredicted in \cite{ochi}
but in \cite{ochi3} the agreement is quite better after the new parametrization.
One of the problems in the fit of \cite{ochi2,ochi3} is that the
range parameter of the $\rho$ coupling to baryons is very small, around 200 MeV, which 
would not be easily accomodated in other areas of the $\rho$ phenomenology, like the
isovector $\pi N$ $s-wave$ scattering amplitude.

The models \cite{ochi,ochi2,ochi3} 
exploit the freedom given by the experimental uncertainties in 
key rates like the partial decay ratios of the $N^\ast(1520)$ resonance. They take 10 $\%$ into $s$-wave $\pi\Delta$, 10 $\%$ into $d$-wave
and 22 $\%$ into the $\rho N$ channel, while the Particle Data Group \cite{pdg}
is offering
bands as: 10-14 $\%$, 5-12 $\%$, 15-25 $\%$, respectively.
Instead of that the model in \cite{tejedor,tejedor2} takes a more conservative
approach and
chooses
the medium value in the Particle Data Group.

Therefore, the model of \cite{tejedor2} has no free parameters. All 
input is obtained uniquely from 
properties of resonances and their decay, with some unknown signs borrowed 
from quark models. 

Another work about these processes is \cite{ripani}. In this paper the authors
extend their predictions to high energy in a
phenomenological way.  They study the photoprodution and electroproduction of $\Delta^{++}\pi^-$
and present results with initial and final state interaction including more high
energy
resonances than in \cite{tejedor,tejedor2,ochi,ochi2,ochi3}.
 However, they are less demanding in questions like gauge invariance, and their
 initial and final state interactions have some ambiguities.

Our aim is to improve the model of 
\cite{tejedor,tejedor2} guided by the new additional experimental results,
trying to find the missing mechanisms in the previous description of the 
$\gamma p\rightarrow\pi^+\pi^0 n$ reaction which bring agreement with the new
data and do not spoil the agreement reached in other pion charge channels.

\section{Model for $\gamma N\rightarrow\pi\pi N$}
\subsection{Brief summary of G\'omez Tejedor-Oset Model}

The model \cite{tejedor,tejedor2} describes double pion photoproduction 
based
on a set of tree level diagrams. These Feynmam diagrams involve pions,
nucleons and nucleonic resonances. Several baryon resonances are included in the
model. They are: $\Delta$(1232) or $P_{33}$ ($J^{\pi}=3/2^+$, I=3/2), 
$N^\ast$(1440) or $P_{11}$ ($J^{\pi}=1/2^+$, I=1/2) and 
$N^\ast$(1520) or $D_{13}$ ($J^{\pi}=3/2^-$, I=1/2). The contribution of the
$N^\ast(1440)$  is small but it was included due to the important role played by
that resonance in the $\pi N\rightarrow\pi\pi N$ reaction and the fact that the
excitation of the $N^\ast(1440)$ peaks around 600 MeV photon energy in the
$\gamma N$ scattering.
The $N^\ast(1520)$ has a large coupling to the photons and is an important
ingredient due to its interference with the dominant term of the process, the
$\gamma N\rightarrow\Delta\pi$ transition called the $\Delta$ Kroll Ruderman contact
term. No other resonances were considered at that time assuming their
contribution to the process would be small in the Mainz range of energies below
$E_\gamma=800 MeV$. Indeed, simple estimates based on the coupling to the
photons of these resonances and their posterior decay into $N\pi\pi$ show that
this would be the case provided there is no interference of terms, which is the
most common possibility given the large freedom in the dependence of the
amplitudes in the momenta and spin of the three particles of the final state.

The Feynman diagrams taken into account are shown in the fig. 1. The amplitudes
are evaluated from the interaction Lagrangians which are shown in the
Appendix A1. The Feynman rules are also shown in the Appendix A2. From there the
amplitudes will be evaluated and they can be found in the Appendix A3, together
with the coupling constants and form factors.

\begin{figure}[h]
\centerline{\protect
\hbox{
\psfig{file=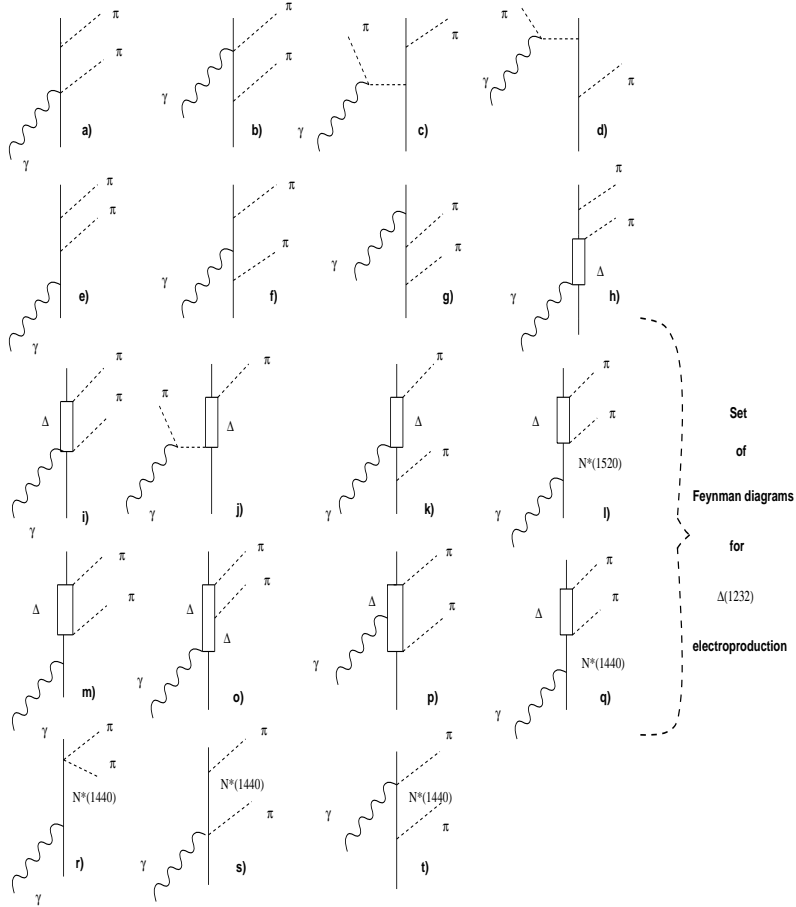,height=12.0cm,width=10.5cm,angle=-90}}}
\caption{\small{Feynman diagrams used in the model for $\gamma
N\rightarrow\pi\pi N$ of ref. \cite{tejedor2}.}}
\end{figure}

The model makes predictions for the six possible isospin channels.
Next we discuss  the relevant interference mechanisms in the $\gamma
p\rightarrow\pi^+\pi^- p$ channel.

The $\Delta$(1232)-intermediate states are dominant in the cross sections.
This dominance comes especially from the $\Delta$ Kroll Ruderman and
$\Delta$-pion pole terms (diagrams $(i)$ and $(j)$). The non resonant terms give
a small background. The contribution coming from the $N^\ast(1520)$ (diagram
$(l)$) is small by
itself if we compare it to the $\Delta$ Kroll Ruderman, but it is crucial to
reproduce the total cross section due to its interference with the $\Delta$
Kroll Ruderman diagram. That interference is responsible for the maximum in the
cross section and is essential for a good agreement with the experimental 
results.
Only the $s$-wave part of the $N^\ast(1520)\Delta\pi$ contribution and the
$\Delta$ Kroll Ruderman are producing that interference. We can see in the results of
\cite{tejedor,tejedor2} that for photon energies below 760 MeV (below the
$N^\ast(1520)$ resonance pole) the interference of the real parts of the
amplitudes $-iT_{(i)}$ and $-iT_{(l)}^{s-wave}$ is constructive while for
energies above the resonance pole it is destructive. That situation plus the
imaginary contribution from $-iT_{(l)}^{s-wave}$ leads to a peak in the cross
section. This interference mechanism appears  in other isospin channels but its
influence is smaller due to isospin coefficients in some cases or due to the fact that in
channels as $\gamma p\rightarrow \pi^0\pi^0 p$ the $\Delta$-Kroll Ruderman term is
zero. 

The status of the results for the different channels in the two pion
photoproduction reactions on the proton is the following:
 
\begin{itemize}
\item
For the $\gamma
p\rightarrow\pi^+\pi^- p$ channel the model reproduces quite well the total
cross sections and the shape of invariant masses up to 800 MeV in photon energy
\cite{bra}.
\item 
For the $\gamma
p\rightarrow\pi^+\pi^0 n$ channel the model fails clearly and understimates the
experimental results in at least 40 $\%$. The theoretical cross section is
smaller due to smaller isospin coefficients in the $\Delta$-Kroll Ruderman term.

\item
Finally, in the case of the $\gamma p\rightarrow\pi^0\pi^0 p$ channel the 
model shows a
good agreement with the newest experimental data for the cross sections and 
invariant masses of ($\pi^0\pi^0$) and ($\pi^0 p$) \cite{wolf}.
\end{itemize}

\subsection{Improvements to the model}

 A model for $\Delta\pi$ electroproduction on the
proton was presented in \cite{nacher}. The aim of this work was to extend the model of
\cite{tejedor,tejedor2} for the $\gamma N\rightarrow\pi\pi N$ 
reaction to virtual photons
selecting the diagrams which have a $\Delta$ in the final state
\footnote{In fig. 1 we also show the diagrams which are included in the $\Delta(1232)$
electroproduction model.}. The agreement found with
$\gamma_v p\rightarrow\Delta^{++}\pi^-$ was good. This 
reaction selecting the $\Delta$ 
final state was an interesting test for the forthcoming full model of the
$\gamma_v N\rightarrow\pi\pi N$ reactions \cite{tesina,nacher2}. 

 We must note that the formalism followed in \cite{nacher} for the vertices of
 the $N^\ast(1520)$ is different from that of \cite{tejedor,tejedor2}. In ref.
 \cite{nacher}
 we followed the paper from Devenish et al.,
\cite{devenish} and we wrote the relativistic current as:
\begin{equation}
J_{e.m.}^{\mu} = G_1(q^2)J_1^{\mu}+
G_2(q^2) J_2^{\mu}+
G_3(q^2) J_3^{\mu}\, ,
\end{equation}
where
\begin{equation}
J_1^{\mu}=\bar{u}_\beta(q^\beta\gamma^\mu -q \!\!\!\!\, / g^{\beta\mu})u\, ,
\end{equation}
\begin{equation}
J_2^{\mu}=\bar{u}_\beta(q^\beta p^{\prime\mu} -p^\prime\cdot q
g^{\beta\mu})u\, ,
\end{equation}
\begin{equation}
J_3^{\mu}=\bar{u}_\beta(q^\beta q^\mu - q^2g^{\beta\mu})u\, ,
\end{equation}
with $G_1$,$G_2$,$G_3$ the electromagnetic form factors for this vertex
and $p^\prime$  the momentum of the resonance.

Taking a non relativistic reduction and using 
$u_\mu$ Rarita-Schwinger spinors
in the c.m. of the resonance, the vertex takes an expression given by:
\\

Scalar part:
\begin{equation}
V^0_{\gamma N N^{\prime\ast}}=i(G_1+G_2 p^{\prime 0} + G_3 q^0)\vec{S}^\dagger
\cdot\vec{q}\, ,
\end{equation}
and the vector part:
 
\begin{eqnarray}
V^i_{\gamma N N^{\prime\ast}}=-i
[(\frac{G_1}{2m} - G_3)(\vec{S}^\dagger\cdot\vec{q})\, \vec{q} - 
\\ \nonumber & & 
\hspace{-5cm} 
iG_1\frac{\vec{S}^\dagger\cdot\vec{q}}{2m}(\vec{\sigma}\times\vec{q})
-\vec{S}^\dagger \{G_1(q^0+\frac{\vec{q}\, ^2}{2m}) 
+ G_2 p^{\prime 0} q^0 + G_3
q^2\}]\, .
\end{eqnarray}

In the case of real photons that we are now interested in, only the vector part is
relevant and the form factors are only a constant in the photon point, determined
by the $A_{3/2}^{N^\ast}$ and $A_{1/2}^{N^\ast}$ transversal helicity
amplitudes at $q^2=0$, with $q^\mu$ the momentum of the photon\footnote{$V^i_{\gamma N N^{\prime\ast}}$ is defined including the complex
factor $-i$}. 

In Refs. \cite{tejedor2,cano}, two solutions for the
coupling of the $N^\ast(1520)$ to the $\Delta$ in $s$ and $d- waves$, 
differring only in a global sign,
 were found
from the respective decay widths. Only a sign was compatible with
the experimental $(\gamma,\pi\pi)$ data, because of the strong interference
between the $\gamma N\rightarrow N^\ast(1520)\rightarrow\Delta\pi$ term
and the $\Delta$ Kroll Rudermam one. Here, in the new formalism the amplitude
of the $\gamma N\rightarrow N^\ast(1520)$ transition changes sign
and consequently the signs of the former $N^\ast(1520)\rightarrow\Delta\pi$
couplings must be changed as we explain below in detail.

For the $N^*(1520) \Delta \pi$ coupling, the simplest Lagrangian allowed by
conservation laws is given by \cite{tejedor}:
\begin{equation}                    \label{simple2}
{\cal L}_{N'^* \Delta \pi} =
i \tilde{f}_{N'^* \Delta \pi} \overline{\Psi}_{N'^*}  \phi^{\lambda}
  T^{\dagger\lambda} \Psi_{\Delta}
\: + \: h.c.\, ,
\end{equation}
\noindent
where ${\Psi}_{N'^*}$, $\phi^{\lambda}$ and $\Psi_{\Delta}$ stand for the
$N^*(1520)$, pion and $\Delta(1232)$ field respectively,
$\, T^{\lambda}$ is the
$1/2$ to $3/2$ isospin transition operator. However, such a Lagrangian only gives rise to $s$-wave $N^*(1520) \rightarrow
\Delta \pi$
decay, there is  a large fraction
of decay into $d$-wave too \cite{pdg,manley}. Furthermore, the
amplitude of Eq. (\ref{simple2}) provides a spin independent
amplitude, while non relativistic constituent quark models (NRCQM) give
a clear spin dependence in the amplitude \cite{cap,cano2}.
We take here for this coupling the following Lagrangian, which,
as shown in \cite{cano}, is supported both by the experiment and the NRCQM.
The Lagrangian is given by
\begin{equation}                \label{ln'dp}
{\cal L}_{N'^* \Delta \pi} =
i  \overline{\Psi}_{N'^*}
\left(
   \tilde{f}_{N'^* \Delta \pi} \: - \:
   \frac{\tilde{g}_{N'^* \Delta \pi}}{{\mu}^2}
   S_i^{\dagger} \partial_i \, S_j \partial_j
\right)
\phi^{\lambda} T^{\dagger\lambda} {\Psi}_{\Delta}
\: + \: h.c.\, ,
\end{equation}
\noindent
with $\mu$ the pion mass.

This Lagrangian gives us the vertex contribution to the $N^*(1520)$ decay
into $\Delta \pi$:
\begin{equation}                \label{hn'dp}
V_{N'^* \Delta \pi} = -
\left(
  \tilde{f}_{N'^* \Delta \pi} \: + \:
  \frac{\tilde{g}_{N'^* \Delta \pi}}{\mu^2}
  \vec{S}^{\dagger} \cdot \vec{k}  \, \vec{S} \cdot \vec{k}
\right)
T^{\dagger\lambda}\, ,
\end{equation}
\noindent
where $\vec{k}$ is the pion momentum.
In order to fit the coupling constants $\tilde{f}_{N'^* \Delta \pi}$
and $\tilde{g}_{N'^* \Delta \pi}$ to the
experimental amplitudes in $s$- and
$d$-wave \cite{pdg} we make a partial wave expansion \cite{pin} of the
transition amplitude $N^*(1520)$ to $\Delta \pi$ from a state of spin 3/2
and third component $M$, to a state of spin $3/2$ and third component
$M'$ and we write it as: 
$$
 \, \langle \frac{3}{2} M' |  V_{N'^* \Delta \pi}
                              | \frac{3}{2}, M \rangle =
A_s C(\frac{3}{2}, 0, \frac{3}{2}; M, 0, M')\, Y_0^{M'-M} 
(\theta, \phi) \: + \:
$$
\begin{equation}                       \label{pwa}
A_d \, C(\frac{3}{2}, 2, \frac{3}{2}; M, M'-M, M')\, 
Y_2^{M'-M}(\theta , \phi)\, ,
\end{equation}
\noindent
where $C(j_1, j_2, J; m_1, m_2 , M)$ is the corresponding
Clebsch-Gordan coefficient, $Y_l^m(\theta, \phi)$ are the spherical
harmonics, and $A_s$ and $A_d$ are the $s$- and $d$-wave partial amplitudes
for the $N^*(1520)$ decay into $\Delta(1232)$ and $\pi$, which are given by:
\begin{equation}            \label{AsAd}
\begin{array}{l}
A_s = - \sqrt{4 \pi}
     \left( \tilde{f}_{N'^* \Delta \pi} \: + \:
            \frac{1}{3} \tilde{g}_{N'^* \Delta \pi} \frac{\vec{k}\,^2}{\mu^2}
     \right)\, , \\ \\
A_d = \frac{\sqrt{4 \pi}}{3} \tilde{g}_{N'^* \Delta \pi}
\frac{\vec{k}\,^2}{\mu^2}\, .
\end{array}
\end{equation}
In \cite{cano} the decay width for the $\Delta\pi$ channel
is given  by


\begin{equation}        \label{gammadp}
\Gamma =
 \frac{1}{4 \pi^2} \, \frac{m_{\Delta}}{m_{N'^*}} \, k
 \left( |A_s|^2 + |A_d|^2 \right) 
\theta (m_{N'^*} - m_\Delta - \mu )\, ,
\end{equation}
\noindent
where $k$ is the momentum of the pion. This expression assumes the $\Delta$
resonance as a stable particle with zero width. In the present work we improve
upon this approximation by explicitly including the $\Delta$ mass distribution
due to the finite width of the $\Delta$ and we have

\begin{eqnarray}    \label{gammadp2}
\Gamma=
\frac{1}{2\pi^2}\int d M_I\frac{\Gamma(M_I)}{(M_I-m_\Delta)^2 +
(\frac{\Gamma(M_I)}
{2})^2}\frac{M_I}{m_{N^{\prime\ast}}}\frac{k(M_I)}{4\pi} 
\left( |A_s|^2 + |A_d|^2 \right)
\\ \nonumber & &
\hspace{-12cm}
\times\theta(m_{N^{\prime\ast}} - M_I -\mu)\, , 
\end{eqnarray}

\noindent 
where $k(M_I)$ = $\frac{\lambda^{1/2}(m_{N^{\prime\ast}}^2, M_I^2, m_\pi^2)}
{2 m_{N^{\prime\ast}}}$ is the pion momentum. 
 We then fit the $s$- and $d$-wave
parts of $\Gamma$ to the average
experimental values \cite{pdg} by keeping the
ratio $A_s/A_d$ positive as deduced from the experimental analysis of the
$\pi N \rightarrow \pi \pi N$ reaction \cite{manley}. We get then
two different solutions which differ only
in a global sign,
\begin{equation}                   \label{2sol}
\begin{array}{rll}
{\it (a)} & \hspace{1cm} \tilde{f}_{N'^* \Delta \pi} = 1.061  &
            \hspace{1cm} \tilde{g}_{N'^* \Delta \pi} =-0.640\, , \\
{\it (b)} & \hspace{1cm} \tilde{f}_{N'^* \Delta \pi} =-1.061  &
            \hspace{1cm} \tilde{g}_{N'^* \Delta \pi} = 0.640\, . \\
\end{array}
\end{equation}

Now, the $\gamma p \rightarrow \pi^+ \pi^- p$ reaction allows us to
distinguish between both solutions, hence providing the relative sign with
respect to the $N^\ast(1520) \rightarrow \gamma N$ amplitude. In our case with the
new formalism the good solution is the $(b)$ option, thus differring in sign 
and 
absolute
value from the results given in \cite{cano}.

Finally we must say that for the width of the $N^\ast(1520)$ in the propagator
we have taken the
explicit decay into the dominant channels ($N \pi$, $\Delta \pi$, $N \rho$)
as it is done in
\cite{cano}
with their energy dependence, improving on the results of \cite{tejedor,tejedor2} where the
energy dependence was taken from the $N \pi$ channel.

Because of the $N^*(1520)$ is a $d$-wave resonance, the energy dependence
of the decay width into $N \pi$ is given by

\begin{equation}
\Gamma_{N'^* \rightarrow N \pi} (\sqrt{s}) =
\Gamma_{N'^* \rightarrow N \pi} (m_{N'^*})
\frac{q^5_{c.m.}(\sqrt{s})}{q^5_{c.m.}(m_{N'^*})}
\, \theta (\sqrt{s} - m -\mu)\, ,
\end{equation}

\noindent
where
$\Gamma_{N'^* \rightarrow N \pi} (m_{N'^*})= 66$ $MeV$ \cite{pdg},
$q_{c.m.}(m_{N'^*})=456$ $MeV$ and
$q_{c.m.}(\sqrt{s})$ is the momentum of the decay pion in the $N^*(1520)$
rest frame.

For the $\Delta \pi$ channel, the energy dependence of the decay width 
is given by Eq. (\ref{gammadp2}).

Finally, for the $N^*(1520)$ decay into $N \pi \pi$ through the $N \rho$
channel is given by

\begin{eqnarray} \label{gammag}
\Gamma_{N'^* \rightarrow N \rho [\pi \pi]} =
\frac{3\, m}{6(2 \pi)^3}
\frac{m_{N'^*}}{\sqrt s}
g_\rho^2  f_\rho^2
\int d \omega_1 d \omega_2 
| D_\rho(q_1 + q_2) | ^2
( \vec q_1 - \vec q_2 \, )^2
\\ \nonumber & & 
\hspace{-10.5cm}\times \theta(1-|A|)\, ,
\end{eqnarray}
with 
\begin{equation}
A=\frac{(\sqrt{s} - \omega_1 - \omega_2)^2 - m - \vec{q}_1\, ^2 -\vec{q}_2\, ^2}
{2 q_1 q_2}\, ,
\end{equation}
where $q_i=(\omega_i ,\vec q_i)$ ($i=1,2$) are the fourmomenta of the
outgoing pions,
$  D_\rho(q_1 + q_2) $ is the $\rho$ propagator including the $\rho$ width,
$f_\rho$ is the $\rho \pi
\pi$ coupling constant ($f_\rho = 6.14$) and $g_\rho$ is the $N'^* N \rho$
coupling constant ($g_\rho=4.52$) 
that we fit from the experimental $N'^* \rightarrow N \rho [\pi \pi]$ decay
width \cite{pdg} \footnote{To avoid confusion we note that the factor 3 in eq.
(16) comes from isospin Clebsch Gordan coefficients when one considers
$N^{\ast(+)}\rightarrow\rho^{+} n + \rho^0 p$, because of the factor
$\vec{\tau}\cdot\vec{\phi}$ in the Lagrangian. In ref. \cite{cano} 
only the total $N^\ast\rightarrow N \rho$ width was needed and 
the present
factor $3\, g_\rho^2$ written there are $g_\rho^2$ with a value for $g_\rho$,
 which was 
$\sqrt{3}$ times bigger than the present one.}.

\section{ $\rho$ meson contribution}
\subsection{$\rho$ production diagrams}

We include two additional mechanism in the model of \cite{tejedor,tejedor2}, 
which were introduced in the approach of ref. \cite{ochi}. In the fig. 2 we can see the Feynman diagrams corresponding to
these terms. The diagram $(a)$ is the diagram which involves the 
$N^\ast(1520)\rightarrow N \rho$ decay mode and it appears in both the channels
$\gamma p\rightarrow \pi^+ \pi^- p$ and $\gamma p\rightarrow \pi^+ \pi^0 n$.
This new diagram is zero for the $\gamma p\rightarrow \pi^0 \pi^0 p$ because the
intermediate $\rho^0$ is not allowed to decay to $\pi^0\pi^0$. 
The set of $\rho$ diagrams chosen here is not gauge invariant by itself.
The $\rho$ pole and nucleon pole terms are needed to preserve gauge invariance.
However, the contribution of these extra terms to the final amplitude is
negligible in the present case.
In the $\rho$ pole case, one of the $\rho$ mesons is very far off-shell, which
makes this term extremely small, something already noticed in ref. \cite{ochi}.
The nucleon pole terms were evaluated in ref. \cite{tejedor} and were also found
negligible.

Taking the Chiral Lagrangian convention for eqs. (42-43-45) in
the Appendix A1, the  vertex for  
the $N^\ast(1520)\rightarrow p \rho^0\rightarrow [\pi^+\pi^-] p$  decay
is written as
\begin{equation}
-iT_{N^\ast\rho^{0} p}=ig_{\rho} f_{\rho} D_{\rho} F_\rho(s_{\rho})
\vec{S}\cdot(\vec{p}_{+}-\vec{p}_{-})
\end{equation}
and for the $N^\ast(1520)\rightarrow n \rho^+\rightarrow [\pi^+\pi^0] n$ is
given by

\begin{equation}
-iT_{N^\ast\rho^{+} n}=-ig_{\rho} f_{\rho} D_{\rho}
 F_\rho(s_{\rho})
\vec{S}\cdot(\vec{p}_{+}-\vec{p}_{0})\sqrt{2}\, ,
\end{equation}
where $D_{\rho}$ is the $\rho$ propagator given by
\begin{equation}
D_{\rho}(q)=\frac{1}{q^2-m^2_{\rho}+ i m_{\rho}\Gamma_{\rho}(\sqrt{s_{\rho}})}
\end{equation}
with $\sqrt{s_{\rho}}=\sqrt{q\,^{0\,2} - \vec{q}\,^2}$ and the $\rho$ decay width given
by
\begin{equation}
\Gamma_{\rho}(\sqrt{s_{\rho}}) = \frac{2}{3} \frac{f_{\rho}^2}{4\pi}\frac{1}{s}
|\vec{p}_{cm}|\, ^3\, . 
\end{equation}
For the off-shell $\rho$ meson we use a form factor $F_\rho$ 
which is shown in the
Appendix A3. Using eq. (\ref{gammag}) we obtain the new value for
${g_\rho} = 5.09$ when we include this form factor in the amplitude. The 
$\sqrt{2}$ factor in eq. (19) is the isospin coefficient from the
$\vec{\tau}\cdot\vec{\phi}_\rho$ coupling.
This isospin factor makes the term $T_{N^\ast\rho N}$ in the
$\gamma
p\rightarrow\pi^+\pi^0 n$ bigger than in the $\gamma
p\rightarrow\pi^+\pi^- p$ reaction. 


The diagram $(b)$ contains a $\gamma N\rho N$ contact interaction or $\rho$ Kroll
Ruderman term. This term will contribute only  to the $\gamma
p\rightarrow\pi^+\pi^0 n$ channel. In the case of 
$\gamma p\rightarrow \pi^+ \pi^- p$, the intermediate $\rho$ meson is neutral
and does not couple to photons. The Feynman rule for the $\gamma N \rho N$ contact term  is 
written as

\begin{equation}
V_{\gamma N \rho N}= e \frac{f_{\rho N
N}}{m_{\rho}}\sqrt{2}(\vec{\sigma}\times\vec{\epsilon}_\gamma)
\cdot\vec{\epsilon}_{\rho}\, , 
\end{equation}
which comes from the $N N\rho$ vertex by minimal substitution. The amplitude for 
diagram  $(b)$, which includes the $\rho$ decay to two pions, is given
by
\begin{equation}
-iT^{\rho}_{KR}=-e \sqrt{2} f_{\rho} 
\frac{f_{\rho N N}}{m_{\rho}} D_{\rho}F_{\rho}(s_{\rho})\, (\vec{\sigma}
\times\vec{\epsilon}_\gamma)\cdot(\vec{p}_+
-\vec{p}_0)\, .
\end{equation}


In the last equations $\frac{f_{\rho N N}}{m_\rho}$ is written as $\sqrt{C_\rho}\frac{f_{\pi N
N}}{m_\pi}$. The constant is $C_\rho$ = 3.96 when the parameter of the form factor
shown in Appendix A3 is taken as $\Lambda_{\rho}$ = 1.4 GeV.

\begin{figure}[h]
\centerline{\protect
\hbox{
\psfig{file=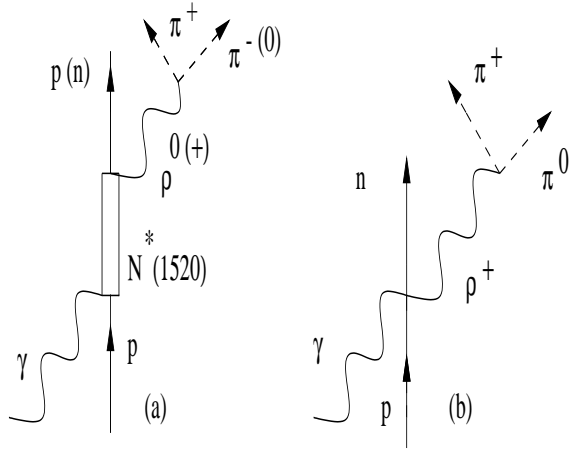,height=6.0cm,width=7.5cm,angle=0}}}
\caption{\small{Feynman diagrams for the $\rho$ meson contribution in the
$\gamma p\rightarrow \pi^+ \pi^- p$ and 
$\gamma p\rightarrow \pi^+ \pi^0 n$ channels. a) The
$N^\ast\rightarrow\rho N$ term. b) The $\rho$ meson Kroll Ruderman term.  }}
\end{figure}

\subsection{Cross sections for $\pi^+\pi^-$ and $\pi^+\pi^0$}

In this section we show the cross sections for the $\gamma
p\rightarrow\pi^+\pi^- p$ and $\gamma p\rightarrow\pi^+\pi^0 n$ including the
$\rho$ meson contribution.
The cross section for the $\gamma N \rightarrow
 \pi \pi N$ reaction
is given by

$\sigma = \frac{m}{\lambda^{1/2}(s, 0, m^{2})}
\frac{S_B}{(2\pi)^{5}}
\int \frac{d^{3} p_{4}}{2\omega_{4}}
\int \frac{d^{3} p_{5}}{2\omega_{5}}
\int d^{3}p_{2}\frac{m}{E_{2}}$

\begin{equation}
\delta^{4} (k + p_{1} - p_{2} - p_{4} -p_{5})
\overline{\sum_{s_i}} \sum_{s_f} |T|^{2}
\end{equation}

$
=\frac{m^{2}}{\lambda^{1/2}(s, 0, m^{2})}
\, \frac{S_B}{4(2\pi)^{4}}
\int d \omega_{5} d \omega_{4} d \cos \theta_{5} d \phi_{45}
$

\begin{equation} \label{total}
\theta (1 - cos^{2} \theta_{45}) \overline{\sum_{s_i}} \sum_{s_f} |T|^{2}\, ,
\end{equation}
where $k = (\omega,\vec{k}\,)$, $p_{1} = (E_{1},\vec{p}_{1})$,
$p_{2} =(E_{2},\vec{p}_{2})$,
$p_{4} = (\omega_{4}, \vec{p}_{4})$, $p_{5} = (\omega_{5},\vec{p}_{5})$
are the momenta
of the photon, incident proton, outgoing proton
and the outgoing pions respectively. $S_B$ is a Bose symmetry factor,
$S_B$ =1/2 for the $\pi^0\pi^0$ final states, and $S_B$=1 otherwise. In Eq. (\ref{total}) 
$\phi_{45}$,
$\theta_{45}$ are the azimuthal and polar angles of $\vec{p}_{4}$ with
respect to $\vec{p}_{5}$ and $\theta_{5}$ is the angle of $\vec{p_{5}}$ with
the $z$ direction defined by the incident photon momentum $\vec{k}$. While $\phi_{45}$ 
is an integration variable, $\theta_{45}$ is given by energy momentum conservation in terms
of the other variables. T is the
invariant matrix element for the reaction.


In fig. 3 we show the contribution of the different terms to the 
$\gamma p \rightarrow\pi^+\pi^0 n$ reaction. We can see
that the $\rho$ Kroll Ruderman term gives by itself a small contribution,
 something already noted in ref. \cite{ochi}. The $N^\ast(1520)$ excitation
followed by $\rho N$ decay shows the $N^\ast$ resonant shape and has a strength
at the peak of about 10 $\mu b$. The $\Delta$ Kroll Ruderman term provides a
background increasing smoothly with the energy. 
In the figure we show the contribution from the coherent sum of the $\Delta$ Kroll Ruderman
and $N^\ast\rho N$ terms in the case of the $g_\rho$ positive (lower solid line) and
with $g_\rho$ negative (lower dark dashed line). We can see that $g_\rho$ positive
leads to constructive interference between these terms, while $g_\rho$ negative
leads to destructive interference. 
However, it is remarkabable that, in spite of the differences between these two latter results, 
when the rest of the contributions are added, the total cross sections are not too different for
either sign of $g_\rho$. This is an interesting observation that calls for justification, but
before we come back to this point we should explain the choice of sign for $g_\rho$.
 Some studies of $\pi N\rightarrow\pi\pi N$ reactions, within the framework of the isobar model,
 extract many resonances parameters \cite{manley,long,vrana}. One of these parameters
 is the relative sign of the decay coupling for the $N^\ast(1520)\rightarrow\rho N$ in $S=3/2$ 
 $s$-wave, 
 to that of $N^\ast(1520)\rightarrow\Delta\pi$,
 which is determined
in those works and leads to positive interference of these two terms at the
$N^\ast(1520)$ energy.


\begin{figure}[h]
\centerline{\protect
\hbox{
\psfig{file=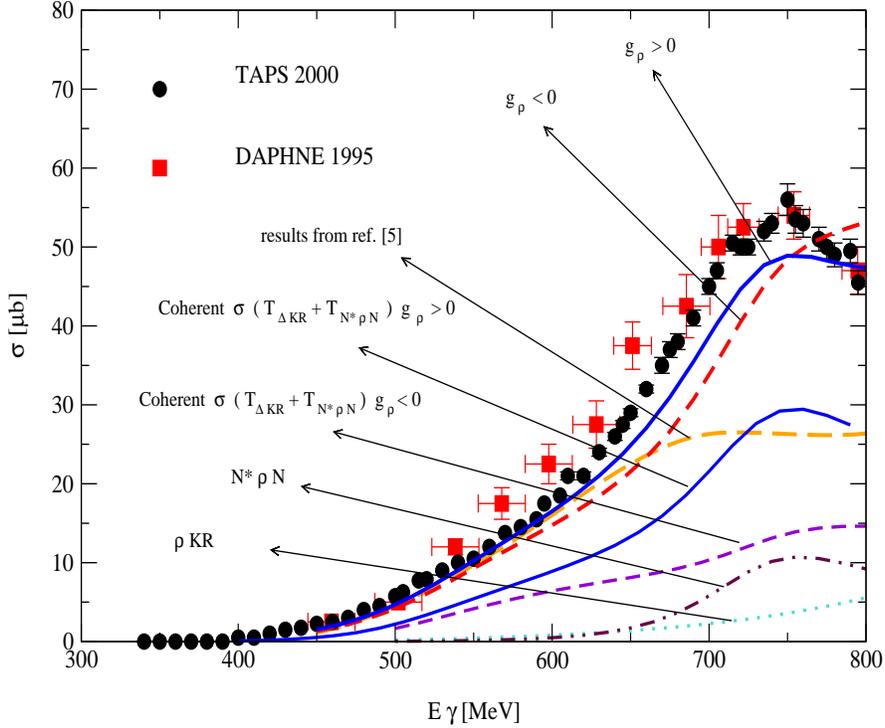,height=12.0cm,width=13.5cm,angle=-90}}}
\caption{\small{ Total cross section for $\gamma
p\rightarrow\pi^+\pi^0 n$: The labels indicate different partial contributions. 
Experimental data from refs. \cite{metag} (circles) and \cite{bra} (squares). }}
\end{figure}

\begin{figure}[h]
\centerline{\protect
\hbox{
\psfig{file=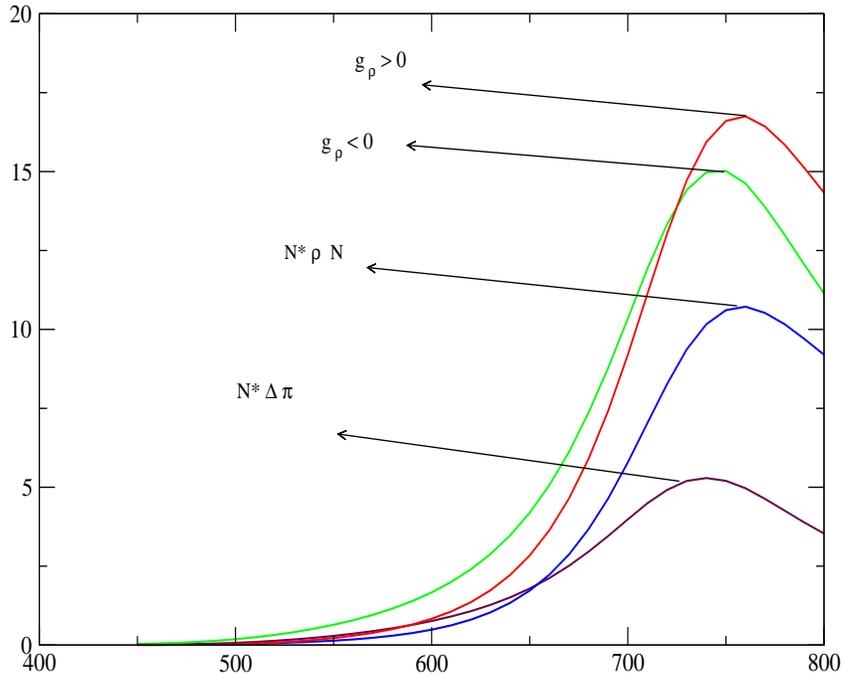,height=12.0cm,width=13.5cm,angle=-90}}}
\caption{\small{ Partial cross sections for $\gamma
p\rightarrow\pi^+\pi^0 n$ with the $N^\ast\Delta \pi$ and $N^\ast\rho N$ terms. The curves
labelled $N^\ast\Delta\pi$ and $N^\ast\rho N$ indicate the contribution of these two
terms. The two upper curves stand for the coherent sum of these two terms with either
sign of $g_\rho$.}}
\end{figure}
In order to see that $g_\rho > 0$ is the right choice we show in fig. 4 that for 
$g_\rho$ positive one gets a larger contribution from the coherent
sum of the $N^\ast\rho N$ and $N^\ast\Delta\pi$ terms than with $g_\rho$ negative
at the $N^\ast(1520)$ pole, indicating constructive interference of the two amplitudes.
It is interesting to note that this sign is the one that leads to better agreement with the
$\gamma p\rightarrow\pi^+\pi^0 n$ data, as shown in fig. 3.

\begin{figure}[h]
\centerline{\protect
\hbox{
\psfig{file=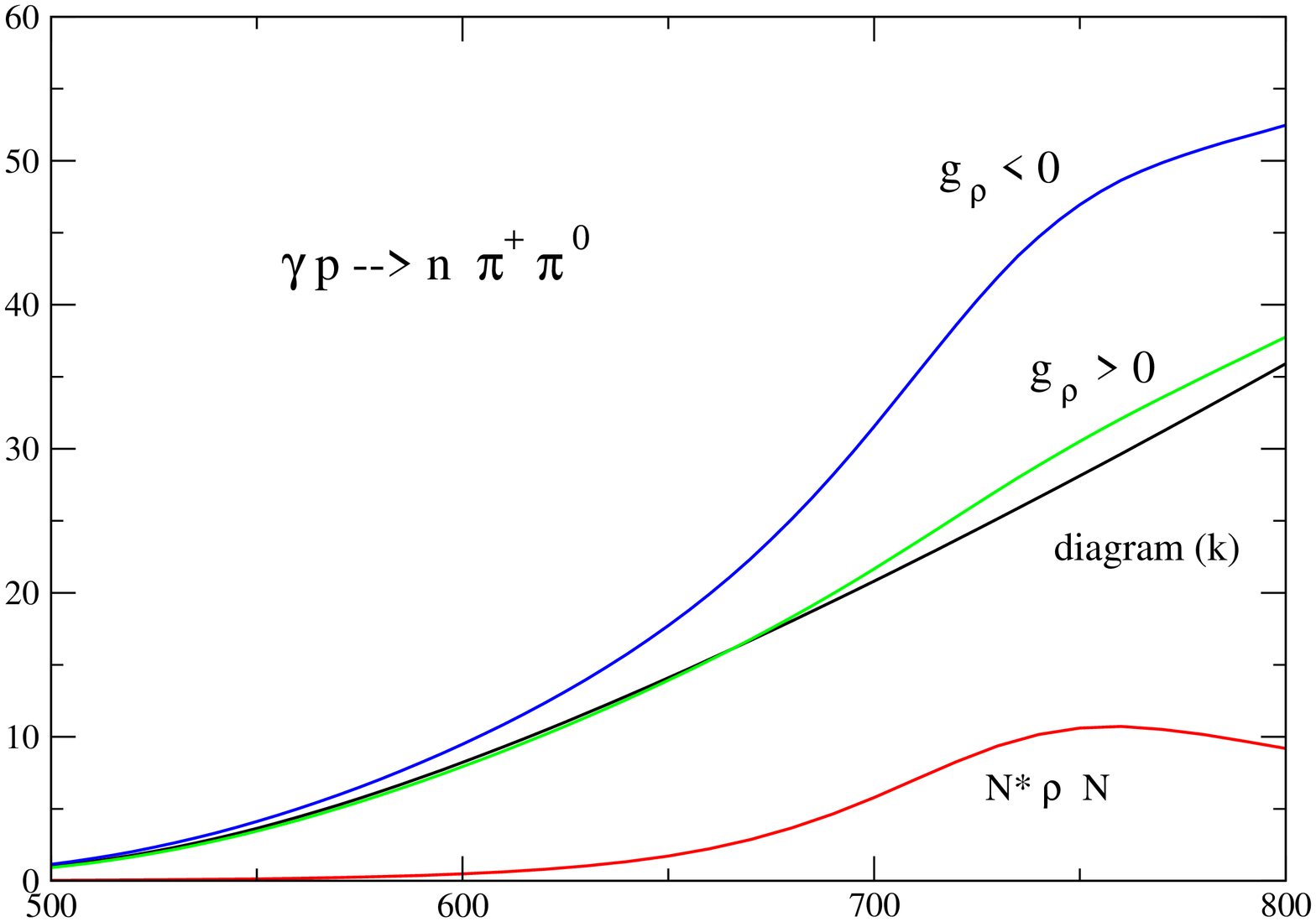,height=12.0cm,width=13.5cm,angle=-90}}}
\caption{\small{ Partial cross sections for $\gamma
p\rightarrow\pi^+\pi^0 n$ with the diagram $(k)$ and $N^\ast\rho N$, 
with the same
meaning as fig. 4.}}
\end{figure}

\begin{figure}[h]
\centerline{\protect
\hbox{
\psfig{file=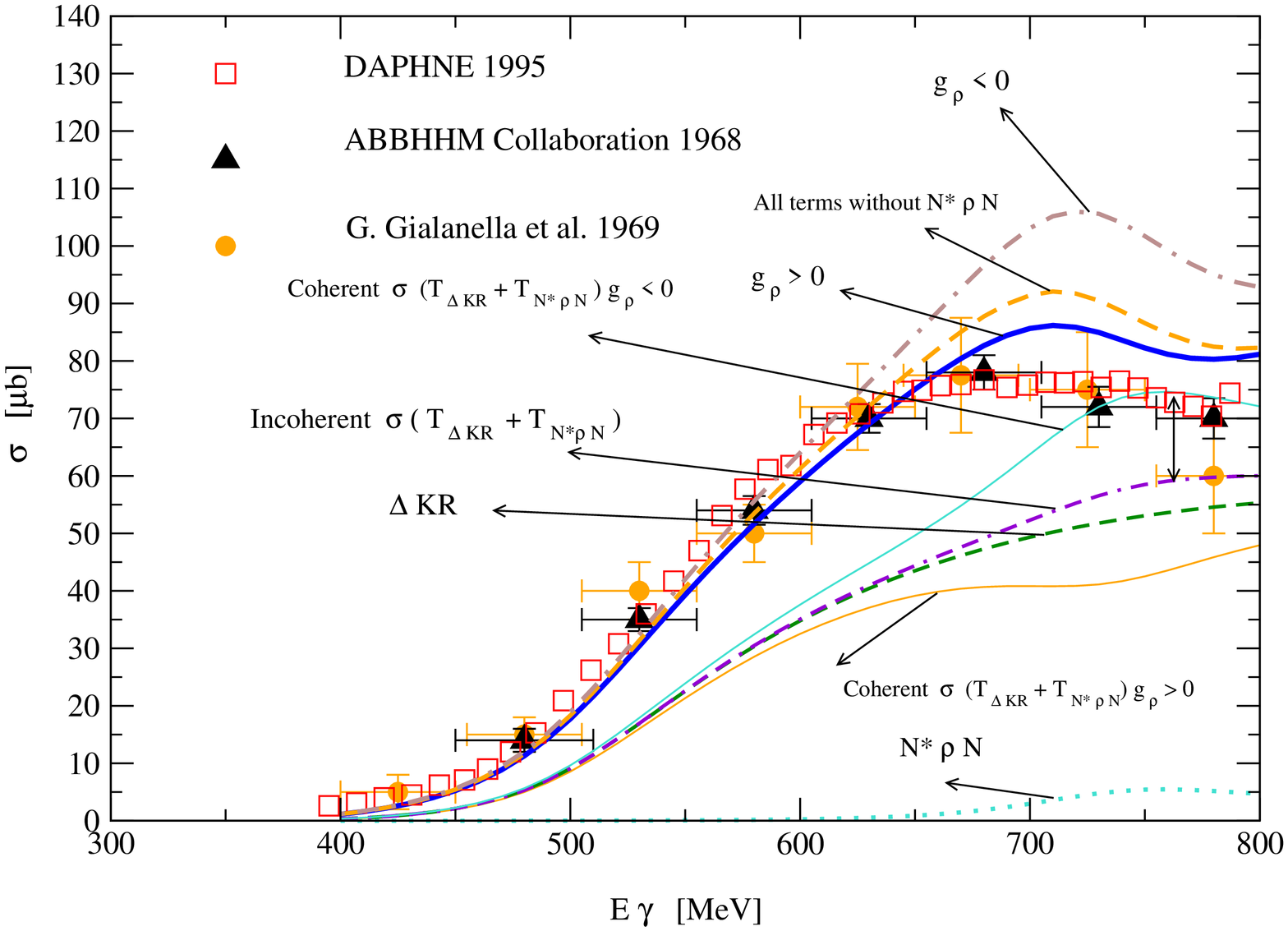,height=12.0cm,width=13.5cm,angle=-90}}}
\caption{\small{Total cross section for $\gamma
p\rightarrow\pi^+\pi^- p$: The labels indicate different partial contributions. 
Experimental data from refs. \cite{bra} (squares), \cite{ABBA} 
(triangles) and \cite{Gial} (circles). }}
\end{figure}

Coming back to fig. 3 and the question of why the total results with the 
two signs of $g_\rho$
are similar although the partial sum of the $\Delta$ Kroll Ruderman and
$N^\ast\rho N$ terms is quite sensitive to this sign, we give
an explanation in fig. 5. As it can be seen there the reason is the large contribution of the diagram $(k)$
of fig. 1, where a pion is emitted prior to the $\Delta$ excitation with the photon. We can see
that this diagram by itself accounts for about one half of the cross section around the 
$N^\ast$ peak. This term is more important here than in the $\gamma p\rightarrow\pi^+\pi^- p$
reaction due to isospin factors (See table A4).
On the other hand there is also an important interference between this term and the 
$N^\ast\rho N$
term but opposite to the one already discussed of the $N^\ast\rho N$ and $\Delta$ Kroll Ruderman
terms. Indeed, we can see in the figure that for $g_\rho < 0$ the interference is constructive
while for $g_\rho > 0$ it is destructive. These two opposite interferences of these importants
terms justify why the final results in fig. 3 were not too different when changing the sign
of $g_\rho$.

In fig. 6 we show the results for the $\gamma p\rightarrow\pi^+\pi^- p$
reaction. In this case there is no $\rho$ Kroll Ruderman term. The 
$\Delta$ Kroll Ruderman term gives a larger strength than in the case
of the $\gamma p \rightarrow\pi^+\pi^0 n$ reaction which is due to isospin
factors. The $N^\ast\rightarrow\rho N$ contribution is in this case a factor two
smaller than in the $\gamma p\rightarrow\pi^+\pi^0 n$ also due to the isospin
coefficients, something already noticed in \cite{ochi}. However, in this
case the interference contribution of the $\Delta$ Kroll Ruderman and
$N^\ast\rightarrow\rho N$ terms is larger than in the 
$\gamma p\rightarrow\pi^+\pi^0 n$ reaction, concretly a factor 3/2, which
can be easily induced, counting again isospin coefficients, due to larger
strength of the $\Delta$ Kroll Ruderman term which more than compensates
the smaller strength of the $N^\ast\rightarrow\rho N$ term.

In the same figure we show the results with the $\Delta$ Kroll Ruderman term alone. We also show
the $N^\ast\rho N$ contribution, which is small by itself. However the interference of this term
with the $\Delta$ Kroll Ruderman term is important, as we already mentioned, and hence the total
cross sections assuming $g_\rho$ positive or negative differ in about 30 $\mu b$. We can also see
that the final results with $g_\rho > 0$, the sign consistent with the isobar model 
analysis, are in much
better agreement with the $\gamma p\rightarrow\pi^+\pi^- p$ data. It is worth mentioning
that the interference between the $\Delta$ Kroll Ruderman and $N^\ast\rho N$ terms is opposite than
in the $\gamma p\rightarrow\pi^+\pi^0 n$ case (see eqs. (18), (19)). It is also instructive
to realize that in the case of $g_\rho > 0$ there is a partial cancellation between the contribution of
the $N^\ast\rho N$ term by itself and the negative interference of this term with the $\Delta$ Kroll
Ruderman one, hence the global effect of the $\rho$ excitation is not as visible here as in the case
of the $\gamma p\rightarrow\pi^+\pi^0 n$ reaction. This will also show up in the $\pi\pi$ invariant
mass distributions that we shall discuss below.


We have already made more comments comparing our results with those of
\cite{ochi}. On the other hand, with respect to the results of ref. \cite{laget}
we find some important differences concerning the strength of the $\rho$ terms,
which in our case and also ref. \cite{ochi} is larger than in \cite{laget}.
In our opinion the main reason for the discrepancies lies in the magnitude of
the $N^\ast(1520)$ photoexcitation. Indeed, the strength found for the
$\gamma p\rightarrow\pi^+\pi^- p$ and $\gamma p\rightarrow\pi^0\pi^0 p$
reactions from the $N^\ast(1520)$ excitation followed by
$N^\ast\rightarrow\Delta\pi$ decay is much smaller than in
\cite{tejedor,tejedor2,cano}. However, it was found experimentally in
\cite{ha} from the analysis of the $N\pi$ invariant mass distributions in
the $\gamma p\rightarrow\pi^0\pi^0 p$ reaction that the main mechanism was the
$N^\ast$ excitation followed by the $\Delta\pi$ decay, supporting the strength
for this mechanism provided by the model of \cite{tejedor,tejedor2,ochi}, and
in strong disagreement with the results of the model of ref. \cite{laget}.

\subsection{Discussion of the interference}

In order to show how this interference appears
we discuss this effect in detail for the $\gamma p\rightarrow\pi^+\pi^- p$ reaction and
we write below the amplitudes of
the diagrams involved in terms of the $\vec{S}$ 
transition spin
operator from $1/2$ to $3/2$. The full amplitudes obtained summing over
intermediate states and operating into initial and final spin states are given
in appendix A4. We have for $\gamma p\rightarrow\pi^+\pi^- p$ with
$\Delta^{++}$ photoproduction\footnote{Full amplitudes with the $D-wave$ part
 are found in appendix A4.}:

\begin{equation} \label{uno}
-i T_{\Delta \, KR}  =  i\frac{f^\ast}{\mu}\vec{S}\cdot\vec{p}_+\,
G_\Delta(\sqrt{s_\Delta})F_\pi((p_--q)^2)\, e\frac{f^\ast}{\mu}
\vec{S}\, ^\dagger\cdot\vec{\epsilon}\, ,
\end{equation} \

\begin{eqnarray}   \label{dos}
-iT_{N^{\prime\ast}\Delta\pi}^{s-wave} & \simeq &
i\frac{f^\ast}{\mu}
\vec{S}\cdot\vec{p}_+\, 
G_\Delta(\sqrt{s_\Delta})G_{N^{\prime\ast}}(\sqrt{s_{N^{\prime\ast}}}) 
 [\tilde{f}_{N^{\prime\ast}\Delta\pi}
+ \frac{1}{3}\frac{\tilde{g}_{N^{\prime\ast}\Delta\pi}}{\mu^2}] 
\\ 
\nonumber
& & 
\times [g_1\, (q^0+\frac{\vec{q}\, ^2}{2m}) + 
g_2\, p^{\prime\, 0} q^0 ]\vec{S}\, ^\dagger\cdot\vec{\epsilon}\, ,
\end{eqnarray}

\begin{equation} 
-i T_{N^{\prime\ast}\rho N}^{s-wave} \simeq  
i\, g_\rho\, f_\rho \vec{S}\cdot(\vec{p}_+ - \vec{p}_-)F_{\rho}
(s_{\rho})
G_{N^{\prime\ast}}(\sqrt{s_{N^{\prime\ast}}})
G_{\rho}(\sqrt{s_{\rho}})\, g_1\, \vec{S}\, ^\dagger\cdot\vec{\epsilon}\, .
\end{equation} 

The coupling constants $g_1$ and $g_2$ come from the photoexcitation
vertex of $N^\ast(1520)$ resonance and their values are given in Appendix
A3.
The interference between the $\Delta$-Kroll Ruderman term and the $s-wave$
$N^\ast(1520)\Delta\pi$ decay contribution ($-i T_{\Delta \, KR}$ and 
$-iT_{N^{\prime\ast}}^{s-wave}$ amplitudes depicted above) was found in 
\cite{tejedor,tejedor2}, and still holds in 
the electroproduction model of \cite{nacher}. As we mentioned in section 2, the
 $-iT_{N^{\prime\ast}\Delta\pi}^{s-wave}$ amplitude has the same structure as
$-i T_{\Delta \, KR}$ and it was seen that for values of the photon energy below 760
MeV ($N^\ast(1520)$ resonance pole)  the interference between the $\it{real}$ part 
of
Eq. (\ref{uno}) and (\ref{dos}) is contructive and for energies above it is destructive.
In the case of the $\rho$ meson contribution the interference also exists but it is
of different nature.  The 
$-i T_{N^{\prime\ast}\rho N}^{s-wave}$ amplitude interferes destructively with the 
$\Delta$ Kroll Ruderman. Indeed, 
both amplitudes summed can be simplified and written as: $-iT^{sum}$ $\sim$
$G_{\Delta}(1+ B\frac{G_{N^{\prime\ast}}}{G_{\Delta}})$, where the propagators 
$G_{\Delta}$ and $G_{N^{\ast\prime}}$
are both mostly imaginary in the region close the pole of the $N^\ast(1520)$.
This is so, in spite of the different masses of the resonances, because in the case of the 
$\Delta$ Kroll Ruderman term the pion emitted prior to
$\Delta$ excitation carries the necessary energy such that the $\Delta$
is left mostly on shell, hence maximizing the strength of the $\Delta$ Kroll Ruderman
mechanism. This interference is
generated between the $\it{imaginary}$ parts of these two amplitudes and the
real parts are not much involved in the effect. This is hence different to the
interference discussed above in eqs. (26, 27), where the interference appeared
in the real parts. Since from eqs. (26, 28) the factor $B$ is negative, 
because $g_\rho > 0$
and the $\rho$ propagator $G_\rho$ is negative, the
interference is destructive.

\section{Other possible high energy resonances}

In this section we discuss the effect of the extra resonances not yet included in the model.
 The additional baryonic
 resonances up to 1.7 GeV are the following \cite{pdg}:

\begin{equation}                            \label{reaction}
\begin{array}{rlcrl}
\hspace{-1cm} {\it (a)} & N (1535,J^{\pi}=1/2^{-}, I=1/2, S_{11}) & \hspace{.5cm} &
\hspace{-1cm} {\it (b)} & N (1650,J^{\pi}=1/2^{-}, I=1/2, S_{11})\\
\hspace{-1cm} {\it (c)} & N (1675,J^{\pi}=5/2^{-}, I=1/2, D_{15}) & \hspace{.5cm} &
\hspace{-1cm} {\it (d)} & N (1680,J^{\pi}=5/2^{+}, I=1/2, F_{15}) \\
\hspace{-1cm} {\it (e)} & N (1700,J^{\pi}=3/2^{-}, I=1/2, D_{13})  & \hspace{.5cm} &
\hspace{-1cm} {\it (f)} & \Delta (1600,J^{\pi}=3/2^{+}, I=3/2, P_{33}) \\
\hspace{-1cm} {\it (g)} & \Delta (1620,J^{\pi}=1/2^{-}, I=3/2, S_{31})  & \hspace{.5cm} &
\hspace{-1cm} {\it (h)} & \Delta (1700,J^{\pi}=3/2^{-}, I=3/2, D_{33})\\
\end{array}
\end{equation}

Most of these resonances can not
appreciably change the results in our range of energies, because their widths
are small and lie at too high energy, because the helicity amplitudes are small,
because the decay width rates into $\Delta\pi$ or $\rho N$ are small or because
a combination of various of these effects:

$(a)-(b)$ The $N^\ast(1535)$ and $N^\ast(1650)$ are $S_{11}$
resonances and their decay modes to $\Delta\pi$ and $\rho N$ are very small.

$(d)$ The $N^\ast(1680)$  resonance is $p$ and $f-wave$ and
the resulting amplitude would not interfere with the interesting terms as
$\Delta$ Kroll Ruderman or $N^\ast(1520)\rightarrow\Delta\pi$.

$(c)-(e)$ The  $N^\ast(1675)$ and $N^\ast(1700)$ have very small photon decay values for both helicity amplitudes
$A_{1/2}$ and $A_{3/2}$. 

$(f)-(g)$ The $\Delta(1600)$ is $p-wave$ and also the helicity amplitudes are very
small and the same small values are found for the $\Delta(1620)$

However, the $\Delta(1700)$ excitation can provide an interesting
contribution. The reasons are the following:
\begin{itemize}
\item Very large Breit-Wigner width. Manley et al., \cite{manley} predicts a
value of 600 MeV. The particle data Table estimates it in around 300 MeV
as average.
\item This $D_{33}$ resonance is $d-wave$ as the $N^\ast(1520)$ ($D_{13}$) and has
similar quantum numbers, hence it should lead to similar interference effects
as found previously for the $D_{13}$ resonance.


\item Its photon decay couplings are large and comparable to those of   
the $N^\ast(1520)$.

\item The $\Delta(1700)$ decay mode into two pion
is very large with $BR(N\pi\pi)$
80-90 $\%$. It includes $30-60$ $\%$ for $\Delta\pi$ decay mode (25-50 $\%$ in 
$s-wave$ and $1-7$ $\%$ in $d-wave$), 30-50 $\%$ into $N\rho$, $S=3/2$
(5-20 $\%$ $s-wave$).

\end{itemize}
Due to these reasons we include that excitation in our model. 
The new diagrams which involve the 
excitation of this resonance can be seen in the fig. 7.
\begin{figure}[h]
\centerline{\protect
\hbox{
\psfig{file=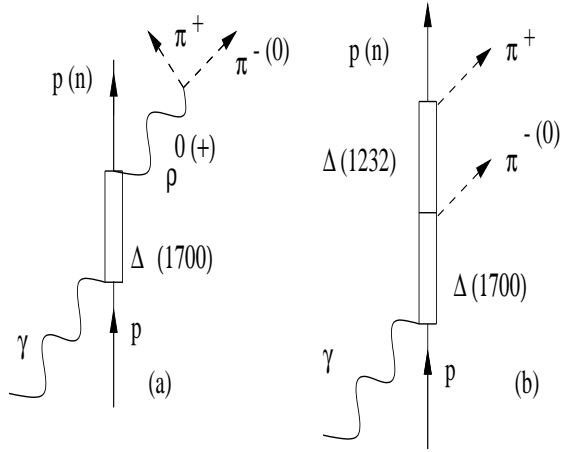,height=6.0cm,width=7.5cm,angle=0}}}
\caption{\small{Feynman diagrams for the $\Delta(1700)$ contribution in the
$\gamma p\rightarrow \pi^+ \pi^- p$ and 
$\gamma p\rightarrow \pi^+ \pi^0 n$ channels. a) The
$\Delta(1700)\rightarrow\rho N$ term. b) The $\Delta(1700)\rightarrow\Delta\pi$
decay amplitude.}}
\end{figure}

The corresponding amplitudes for the diagrams  of the $\gamma p\rightarrow\pi^+\pi^- p$ 
reaction are written as \footnote{ We
shall
express $\Delta(1700)$ as $\Delta^\ast$ in what follows.}:

\begin{eqnarray}
\hspace{-1cm}
-iT_{\Delta(1700)}^{s-wave} & \simeq &
-i\sqrt{\frac{3}{2}}\frac{f^\ast}{\mu}
\vec{S}\cdot\vec{p}_+\, 
G_\Delta(\sqrt{s_\Delta})G_{\Delta^\ast}(\sqrt{s_{\Delta^\ast}}) 
 [\tilde{f}_{\Delta^\ast\Delta\pi}
+ \frac{1}{3}\frac{\tilde{g}_{\Delta^\ast\Delta\pi}}{\mu^2}] 
\\ 
\nonumber
& & 
\times [g_1^\prime\, (q^0+\frac{\vec{q}\, ^2}{2m}) + 
g_2^\prime\, p^{\prime\, 0} q^0 ]\vec{S}\, ^\dagger\cdot\vec{\epsilon}\, ,
\end{eqnarray}

\begin{equation}
\hspace{-1cm}
-i T_{\Delta(1700)\rho N}^{s-wave} \simeq  
i\sqrt{\frac{2}{3}}\, g^\prime_\rho\, f_\rho \vec{S}\cdot(\vec{p}_+ - \vec{p}_-)
F_{\rho}(s_{\rho})
G_{\Delta^\ast}(\sqrt{s_{\Delta^\ast}})
G_{\rho}(\sqrt{s_{\rho}})\, g_1^\prime\, \vec{S}\, ^\dagger\cdot\vec{\epsilon}\,
.
\end{equation} 
The coupling constants $g_1^\prime$ and $g_2^\prime$ coming from the
photoexcitation of the $\Delta(1700)$ resonance are evaluated from the
experimental helicity amplitudes $A_{1/2}$, $A_{3/2}$ following the procedure of
ref. \cite{nacher} and
are given in the Appendix A3.
In eq. (30) the $-\sqrt{3/2}$ factor is an isospin coefficient which accounts
also for the phase of the $\pi^+$,  $|\pi^+\rangle$ =
$- |1 1\rangle$ in isospin base. On the other hand the contribution
$[g_1^\prime\, (q^0+\frac{\vec{q}\, ^2}{2m}) + 
g_2^\prime\, p^{\prime\, 0} q^0 ] $ is proportional to the $A_{3/2}$ amplitude
\cite{nacher}, showing explicitly that the interference with the $\Delta$ Kroll
Ruderman comes from this helicity amplitude, as it was the case in the 
$N^{\ast\prime}$ excitation followed by $\Delta\pi$ decay, eq. (27). Hence
assuming that the $s$-wave part of the $\Delta^\ast\rightarrow\Delta\pi$ decay,
$[\tilde{f}_{\Delta^\ast\Delta\pi}
+ \frac{1}{3}\frac{\tilde{g}_{\Delta^\ast\Delta\pi}}{\mu^2}]$
has the same sign as for the $N^{\ast\prime}\rightarrow\Delta\pi$ decay, the
present term will have destructive interference with the $\Delta$ Kroll Ruderman
term below the pole of the $\Delta^\ast$.
Equations (13,16) were used, taking $\Delta(1700)$ resonance instead
$N^\ast(1520)$, to extract the value of the 
$\tilde{f}_{\Delta^\ast\Delta\pi} = -1.325$,
$\tilde{g}_{\Delta^\ast\Delta\pi} = 0.146$ and $g^\prime_\rho = 2.60$
The relative sign of the $\tilde{f}_{\Delta^\ast\Delta\pi}$ and
$\tilde{g}_{\Delta^\ast\Delta\pi}$ is determined by the relative sign between
the $s$- and $d-$ wave amplitudes in the decay.

The overall sign of these coupling constants  
 is choosen in order to get a destructive
interference. An opposite global sign produces large disagreement with the
experimental results. In the case of the $g_\rho^\prime$ coupling, there is
freedom to choose its sign. The Particle Data Group shows both solutions 
for the relative
sign of the $g_\rho^\prime$ and the couple ($\tilde{f}_{\Delta^\ast\Delta\pi}$ and
$\tilde{g}_{\Delta^\ast\Delta\pi}$) deduced from the experimental analysis of the
$\pi N\rightarrow\pi \pi N$ reaction \cite{manley,long}. The contribution of
this diagram is much smaller than the one in fig. 7$(b)$ and the relevance of choosing 
either sign
is not so important as in the case in the $N^{\ast}\rho N$ diagram
explained before.

\section{Final results}
\subsection{Cross sections for $\pi^+\pi^-$, $\pi^+\pi^0$ and $\pi^0\pi^0$}
In this section we show the final cross section with the effect of $\Delta(1700)$
excitacion plus the effect coming from the $\rho$ meson in the $\gamma
p\rightarrow \pi^+\pi^- p$ and $\gamma p\rightarrow\pi^+\pi^0 n$ reactions. In
the channel $\gamma p\rightarrow  \pi^0\pi^0 p$ the only new diagram which gives
contribution is the $\Delta(1700)$ resonance and its effects are also shown.

\begin{itemize}
\item In the $\gamma p\rightarrow\pi^+\pi^- p$ we have analysed the contribution
coming from the $\Delta(1700)$ and we have discovered that 
by itself the contribution of the diagram is
small even at energies above 700 MeV. However, we have found another interesting
interference effect between the diagram involving the $\Delta$(1700)
excitation and the $\Delta$ Kroll-Ruderman dominant term. In fig. 8, 
we see the contribution coming from the diagram of the $\Delta$(1700)
excitation which is small. However, once interference effects are considered we see a moderate decrease of the total cross
section which would bring our results in better agreement with the results of \cite{bra}.


\begin{figure}[h]
\centerline{\protect
\hbox{
\psfig{file=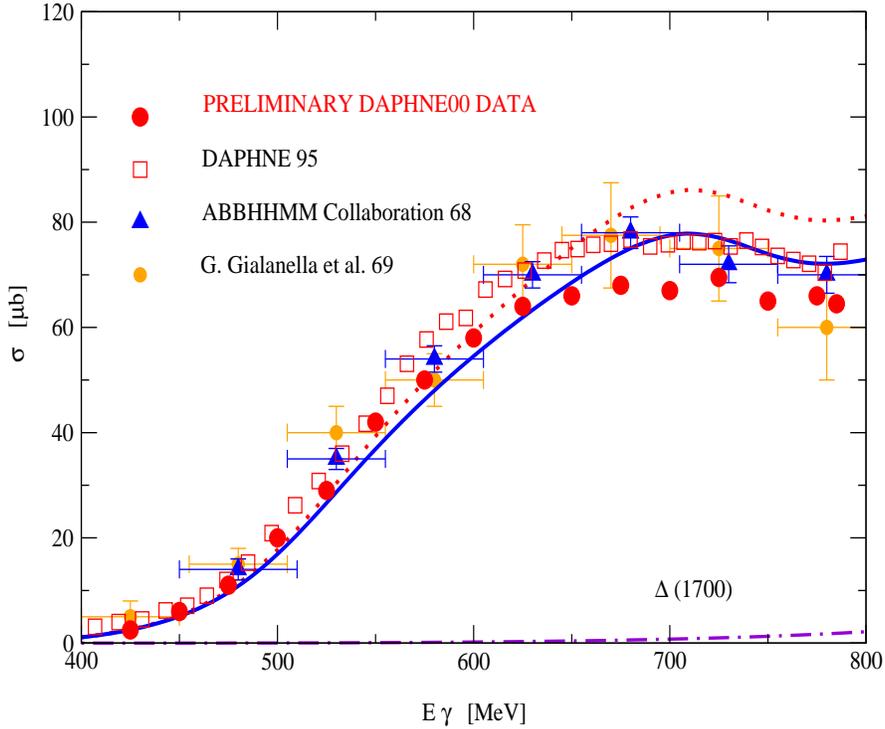,height=12.0cm,width=13.5cm,angle=-90}}}
\caption{\small{Total cross section for $\gamma
p\rightarrow\pi^+\pi^- p$: Continuous line with $\rho$ meson contribution plus $\Delta(1700)$
excitation. Dotted line: cross section without the $\Delta(1700)$
contribution. Dash-dotted line: contribution of the $\Delta(1700)$ terms by
themselves. Experimental data from refs. \cite{bra} (squares), \cite{ABBA} 
(triangles), \cite{Gial} (light circles) and \cite{ahrens} (dark circles).}}
\end{figure}


\item In the case of the $\gamma p\rightarrow  \pi^+\pi^0 n$ we see in fig. 9
that the
effect of this new resonance is less important than in the
$\gamma p\rightarrow\pi^+\pi^- p$ reaction. The reason is that here the diagram
of $\Delta$ Kroll Ruderman diagram 
is less important and the effects of its interference with the $\Delta^\ast$ contribution
are smaller. We can see the final results in the continuous line of fig. 9
 and the contribution coming only from the $\Delta$(1700) is shown in the 
 dashed
line and the results without the $\Delta(1700)$ is dotted line. 

\begin{figure}[h]
\centerline{\protect
\hbox{
\psfig{file=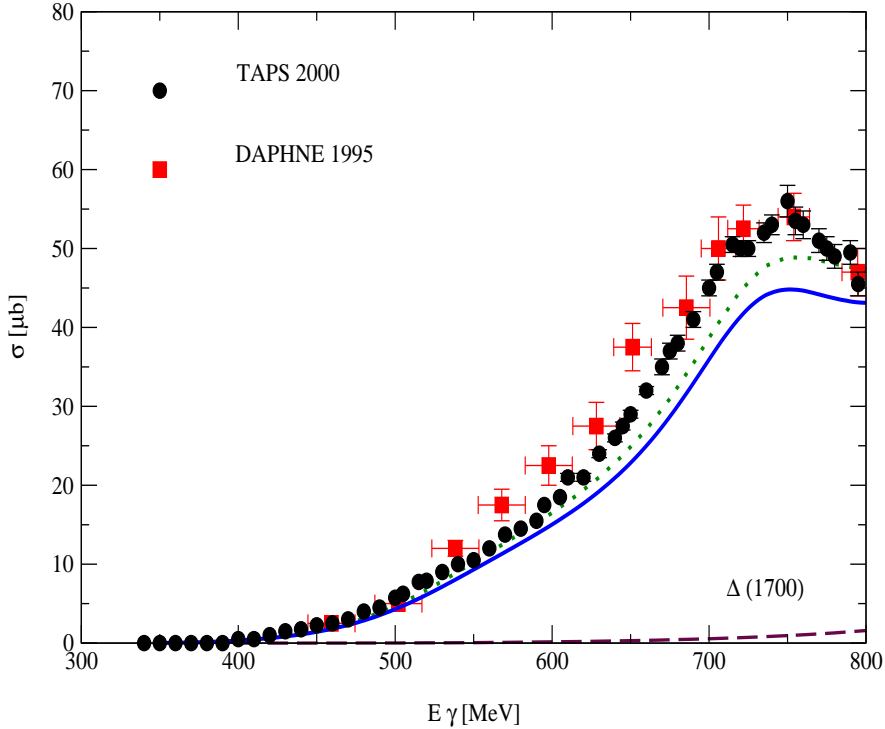,height=12.0cm,width=13.5cm,angle=-90}}}
\caption{\small{Total cross section for $\gamma
p\rightarrow\pi^+\pi^0 n$: Continuous line 
with $\rho$ meson contribution plus $\Delta(1700)$ terms. Dotted line: cross
section without $\Delta(1700)$ terms. Dashed line: contribution from the
$\Delta(1700)$ terms by themselves. Experimental data from refs. \cite{metag}
(circles) and \cite{bra} (squares).}}
\end{figure}

\item The last channel analysed with a proton in the initial state is the
$\gamma p\rightarrow \pi^0\pi^0 p$. The effect of the $\Delta(1700)$ here is also
small. Here the $\Delta$ Kroll Ruderman term is zero since the photon does
no couple to neutral pions. Then the only contribution to this cross section
comes from the diagram alone and not from an interference effect. We show in fig. 10 
the final results with a continuous
line for this channel. The dashed line shows the contribution of the
$\Delta(1700)$ excitation and the dotted line the results of the model without the 
$\Delta(1700)$
resonance. This picture is interesting because it shows clearly that since
the $\Delta$ Kroll Ruderman term is not present, the contribution of the
$\Delta(1700)$ to the cross section is given by the diagram itself without large
interference effects. The agreement of the results with the data is fair, overestimating them a bit in the
peak. This small increase of the results with respect to the old ones in \cite{tejedor2} are due to the new
constants used to $N^\ast(1520)\rightarrow\Delta\pi$ decay when the 
convolution with the width of the
$\Delta$ (eq. (13)) is done. We should also note that the
branching ratio for this decay in PDG table 
\cite{pdg} is 15-25 $\%$ and we take the average value of 20 $\%$, hence, uncertainties tied to this
branching ratio have to be assumed in the theoretical predictions for this channel. Note also that the
contribution from the $N^\ast(1520)$ excitation decaying into $\Delta\pi$ is dominant here
and goes quadratic in the corresponding amplitude, while the influence of this mechanism in the other channels
comes from interference terms which are linear in the $\gamma N\rightarrow N^\ast\rightarrow\Delta\pi$
amplitude. This makes the $\gamma p\rightarrow \pi^0\pi^0 p$ channel more sensitive than the other ones to the
experimental uncertainties in the mentioned branching ratio.
We should also metion that the diagram ($k$) of the Fig. 1 has a sizeable
strength in this channel, but interference with other $\Delta$ background
terms reduces its contribution. 
 
\end{itemize}

\begin{figure}[h]
\centerline{\protect
\hbox{
\psfig{file=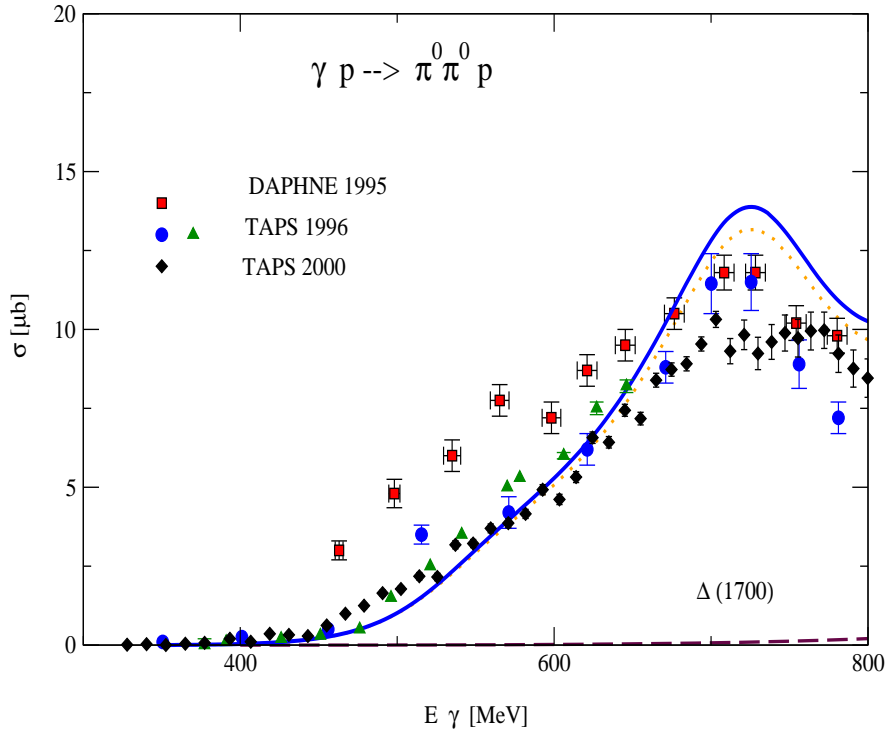,height=12.0cm,width=13.5cm,angle=-90}}}
\caption{\small{Total cross section for $\gamma
p\rightarrow\pi^0\pi^0 p$ with $\Delta(1700)$ contribution with continuous line
and without $\Delta(1700)$ in dotted line. Dashed line: contribution from the
$\Delta(1700)$ excitation terms by themselves. Experimental 
data from refs. \cite{metag} (diamonds), \cite{bra} (squares) and 
\cite{ha} (circles and triangles).}}
\end{figure}

\subsection{Mass distributions} 
We use the
new information about invariant masses of 
$\gamma p \rightarrow \pi^+\pi^0 n$ \cite{metag}
to compare with our improved 
model of two pion photoproduction on the proton at an intermediate range of
photon energies. The recent experimental results about $\gamma
p\rightarrow\pi^0\pi^0 p$ \cite{wolf} are also compared.
 We have seen in our predictions for the cross section in the channel
$\gamma p\rightarrow\pi^+\pi^0 n$ that the
$N^\ast(1520)\rightarrow\rho N$ decay and the $\rho$ Kroll Ruderman terms 
contribute appreciably to the region 
above 600 MeV of photon energy in the total cross section.
We analyse now the invariant mass distributions of ($\pi^+\pi^0$) and 
compare them  with the
experimental results in order to get an additional test of this mechanism.

In fig. 11 we show a set of figures for different bins of photon energies for the
invariant mass of ($\pi^+\pi^0$). The bins are 540-610 MeV, 610-650 MeV, 
650-700 MeV, 700-740 MeV, 740-780 MeV and 780-820 MeV.

\begin{figure}[h]
\centerline{\protect
\hbox{
\psfig{file=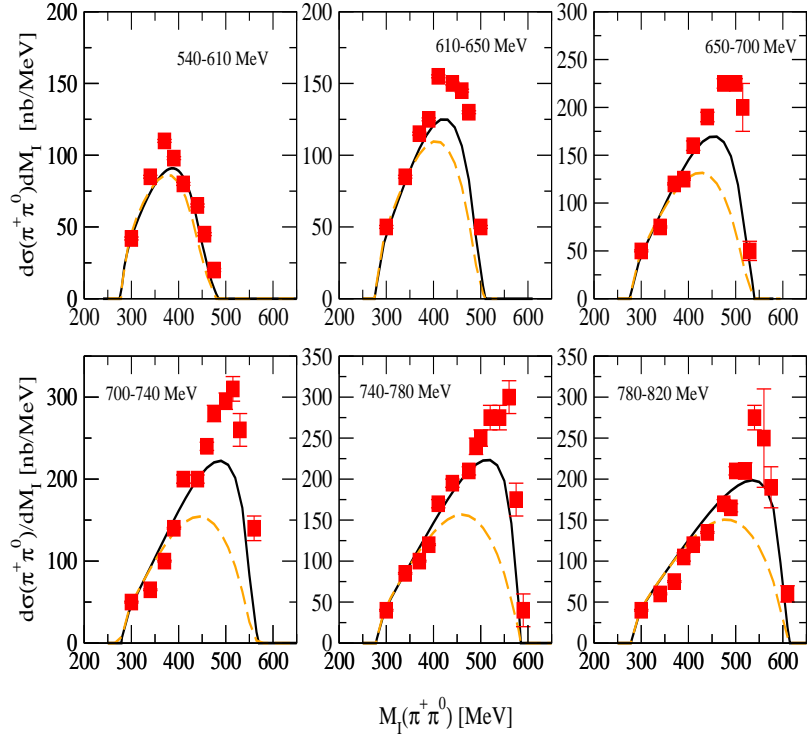,height=12.0cm,width=12.5cm,angle=-90}}}
\caption{\small{Differential cross section with respect to the invariant mass
of the $(\pi^+ \pi^0)$ system for different values of $E_\gamma$ from 540 MeV to 820
MeV for the $\gamma p\rightarrow\pi^+\pi^0 n$ reaction. 
With continuous line we show the
final results with $\rho$ meson and $\Delta(1700)$ terms and with dashed line we
show the results of the model without those contributions. 
Experimental data from ref. \cite{metag}.}}
\end{figure}

We show with a dashed line the results of the model without the new
resonances and with continuous line our final results. 
In the bins of energy above 650 MeV we see the influence of the new terms.
We find an important contribution to the invariant masses due to the tail of the 
$\rho$ meson coming from the diagrams 2$(a)$ and 2$(b)$, moving the strength to
higher energies of the spectrum. These results are
consistent with our predictions for the total cross section and they reassert 
 the
influence of the $\rho$ production mechanisms.

In fig. 12 we show a similar set of figures as we explained before for the 
$\gamma
p\rightarrow\pi^0\pi^0 p$ reaction with the same bins of photon energies.
In continuous line we show the results with the new mechanism of the 
$\Delta(1700)$ diagram
added and with
dotted line the results without that new contribution. We observe
that the new mechanisms have a very small effect in the distributions.
These results are different from those of phase space
alone which are
shown in \cite{wolf}, which implies the existence of interesting structure in that
channel coming esentially from the $N^\ast(1520)\rightarrow\Delta\pi$ decay.

\begin{figure}[h]
\centerline{\protect
\hbox{
\psfig{file=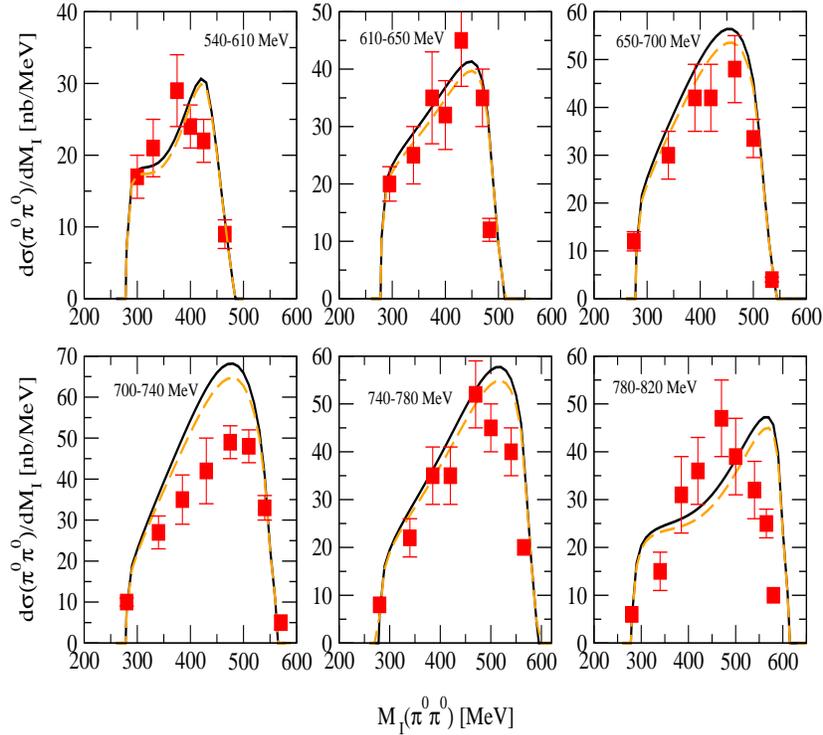,height=12.0cm,width=12.5cm,angle=-90}}}
\caption{\small{Differential cross section with respect to the invariant mass
of the $(\pi^0 \pi^0)$ system for different values of $E_\gamma$ from 540 MeV to 820
MeV for the $\gamma p\rightarrow\pi^0\pi^0 p$ reaction. With continuous line we show the
final results with $\rho$ meson and $\Delta(1700)$ terms and with dashed line we
show the results of the model without those contributions. 
Experimental data from ref. \cite{wolf}.}}
\end{figure}

\begin{figure}[h]
\centerline{\protect
\hbox{
\psfig{file=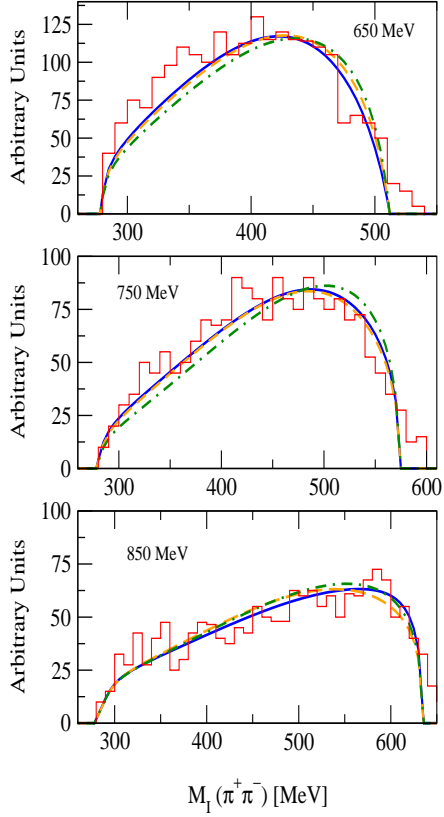,height=12.0cm,width=12.5cm,angle=-90}}}
\caption{\small{Differential cross section with respect to the invariant mass
of $(\pi^+ \pi^-)$ system for different values of $E_\gamma$ for the 
$\gamma p\rightarrow\pi^+\pi^- p$. With continuous line we show the
final result with $\Delta(1700)$ term and with light dashed line we
show the results of the model without this contribution. The dash-dotted line means the
same as the continuous line but with an opposite sign for the $g_\rho$ coupling. The
photon energies are from up to down: 650 MeV, 750 MeV and 850 MeV.
Experimental data from ref. \cite{bra}.}}
\end{figure}

In fig. 13 we show the differential cross section with respect to the invariant
mass of the $(\pi^+ \pi^-)$ system for different values of the photon energy up to
$E_{\gamma}$ = 850 MeV for the  
$\gamma p\rightarrow\pi^+\pi^- p$ reaction. From up to down we show the results
for 650 MeV, 750 MeV and 850 MeV of photon energy.
The experimental data are given in terms of counts, hence the normalization is
arbitrary. We match our results to the peak of the distribution. The model
reproduces the distribution quite well, however we have analysed several cases
and some comments are needed to explain them.
We show three cases in
the figures. The final results with $\rho$ meson and $\Delta(1700)$
terms included are shown with continuous line, 
the results of the model without those contributions are with
light dashed line, and with the dash-dotted line we show the same as
continuous line but with opposite sign for the $g_{\rho}$.
At 650 MeV photon energy all three lines are very close and we observe a
phase space-like distribution. However, as we go to 750 MeV photon energy 
the continuous line is in better agreement
with the data. If we move to even higher energies, at 850 MeV, all three curves
show a peak around the same position, and have similar shapes.  
What we can say from the former discussion is that the spectra 
contains less information on the reaction mechanisms than in the case of the 
$\gamma p\rightarrow\pi^+\pi^0 n$ reaction. Yet, future experiments for those
distributions with the proper normalization should be useful for further test of
the models.

\section{Conclusions}
We have made a new analysis of the $\gamma p\rightarrow\pi \pi N$ reaction
channels. The cross section for $\gamma p\rightarrow\pi^+\pi^- p$,
$\gamma p\rightarrow\pi^+\pi^0 n$ and $\gamma p\rightarrow\pi^0\pi^0 p$ were
calculated with the new additional contributions of $\rho$ meson production and
$\Delta(1700)$ excitation.

The improvements made to the original model \cite{tejedor,tejedor2} were aimed at
obtaining a better understanding of the mechanisms involved in these 
pion photoproduction reactions.

The calculated cross section and invariant masses showed a much better 
agreement with the
experimental results than found in $\cite{tejedor,tejedor2}$ 
for the $\gamma p\rightarrow\pi^+\pi^0 n$. The improvements in the 
$\gamma p\rightarrow\pi^+\pi^0 n$ channel have been done without spoiling the
agreement found previously in the other channels. The 
$\gamma p\rightarrow\pi^0\pi^0 p$ was not much changed by the only new mechanism
which contributes, the $\Delta(1700)$, because the $\Delta$ 
Kroll Ruderman term in this case is absent and there are no interferences. 
Some small changes with respect to previous results are due to
the use of a convolution with the $\Delta$
width in the study of the $N^\ast(1520)\rightarrow\Delta\pi$ decay. However, the final small
changes found in the $\gamma p\rightarrow\pi^+\pi^- p$ channel resulted as a
consequence of partial cancellations between the $N^\ast\rho N$ contribution by itself
and the
destructive interferences of the $\Delta(1700)$ and $N^\ast\rho N$ terms with the $\Delta$
Kroll Ruderman term. The results reported here bring new light to the old
problem of the $\gamma p\rightarrow\pi^+\pi^0 n$ reaction. The new elements
introduced have been stimulated by new experimental measurements that gave clear
indications that the $\rho$ production mechanism was important in that reaction.
 We hope that
the extension of the experiments at higher energies and further theoretical
studies will help unveil new interesting mechanisms and the subtle way that the
resonances influence these reactions.
\\

{\bf Acknowledgements}
\\

We would like to thank M. Wolf, W. Langgaertner, V. Metag and S. Schadmand for
multiple discussions around their preliminary data.
This work has been partially supported by DGICYT
contract number BFM2000-1326.
One of us J.C. Nacher wishes to acknowledge financial support from
the Ministerio de Educaci\'on y Cultura.

\newpage
\begin{center}
APPENDIX
\end{center}

{\Large{\bf A1. Lagrangians.}}

\begin{equation}
 L_{\pi N N} = - \frac{f}{\mu}\overline{\Psi}\gamma^\mu\gamma_5
\partial_\mu\vec{\phi}\cdot\vec{\tau}\Psi
\end{equation}
\begin{equation}
 L_{\Delta \pi N} = - \frac{f^\ast}{\mu}\Psi^\dagger_\Delta S^\dagger_i(
\partial_i\phi^\lambda)T^{\lambda\dagger}\Psi_N  +  h.c.
\end{equation}
\begin{equation}
 L_{\Delta \Delta \pi} = - \frac{f_\Delta}{\mu}\Psi^\dagger_\Delta 
S_{\Delta i}(\partial_i\phi^\lambda)T^\lambda_\Delta\Psi_\Delta  +  h.c.
\end{equation}
\begin{equation}
 L_{N^{\ast} \Delta \pi} = - \frac{g_{{N^\ast}\Delta\pi}}{\mu}
\Psi^\dagger_\Delta  
S^\dagger_i(\partial_i\phi^\lambda)T^{\lambda\dagger}\Psi_{N^\ast}  +  h.c.
\end{equation}
\begin{equation}
 L_{N^{\ast\prime} \Delta \pi} = i \overline{\Psi}_{N^{\ast\prime}}(
\tilde{f}_{N^{\ast\prime} \Delta \pi}-\frac{\tilde{g}_{N^{\ast\prime} 
\Delta \pi}}
{\mu^2}S_i^\dagger\partial_i S_j\partial_j)\phi^\lambda T^{\lambda\dagger}
\Psi_\Delta  +  h.c.
\end{equation}
\begin{equation}
 L_{N N \gamma} = - e\overline{\Psi}_N(\gamma^\mu A_\mu -\frac{\chi_N}{2m}
\sigma^{\mu\nu}\partial_\nu A_\mu)\Psi_N
\end{equation}
\begin{equation}
 L_{\pi\pi\gamma} = ie(\phi_+\partial^\mu\phi_- - \phi_-\partial^\mu\phi_+)A_\mu
\end{equation}
\begin{equation}
 L_{N^\ast \pi N} = - \frac{\tilde{f}}{\mu}\Psi^\dagger_{N^\ast}\sigma_i
(\partial_i\vec{\phi})\cdot\vec{\tau}\Psi_N  + h.c.
\end{equation}
\begin{equation}
 L_{N^\ast \pi \pi N } = - \tilde{C}\overline{\Psi}_{N^\ast}
\vec{\phi}\cdot\vec{\phi}\Psi_N  + h.c.
\end{equation}
\begin{equation}
L_{\gamma\pi N N} = - iq_\pi\frac{f}{\mu}\overline{\Psi}\gamma^\mu\gamma_5
A_\mu\vec{\phi}\cdot\vec{\tau}\Psi
\end{equation}


\begin{equation} \label{l1}
L_{\rho\pi\pi} =
f_{\rho}\vec{\phi}_\mu^{(\rho)}\cdot(\vec{\phi}\times\partial^\mu\vec{\phi})
\end{equation}

\begin{equation}  \label{l2}
L_{N^{\ast\prime} N \rho} = -g_{\rho}\overline{\Psi}_N\,
S_i\vec{\phi}^{(\rho)}_i\cdot\vec{\tau}\Psi_{N^\prime\ast} + h.c.
\end{equation}

\begin{equation}
 L_{\Delta^{\ast} \Delta \pi} = i \overline{\Psi}_{\Delta^{\ast}}(
\tilde{f}_{\Delta^{\ast} \Delta \pi}-\frac{\tilde{g}_{\Delta^{\ast} 
\Delta \pi}}
{\mu^2}S_i^\dagger\partial_i S_j\partial_j)\phi^\lambda T_{\Delta}^{\lambda\dagger}
\Psi_\Delta  +  h.c.
\end{equation}

\begin{equation}  \label{l22}
L_{\Delta^{\ast} N \rho} = -g_{\rho}\overline{\Psi}_N\,
S_i\vec{\phi}^{(\rho)}_i\cdot\vec{T}\Psi_{\Delta^\ast} + h.c.
\end{equation}

Instead of writing the explicit expressions for the terms involving the 
photon and the
excitation of resonances like $L_{\Delta N \gamma}$, $L_{N^\ast N \gamma}$,
 $L_{\Delta\pi\gamma N}$, $L_{N^\ast\pi\gamma N}$, $L_{N^{\ast\prime} N \gamma}$, 
 $L_{\Delta^\ast \Delta \gamma}$
 we
 address the reader directly to the corresponding Feynman rules in the Appendix
 A2, which provide the vertex function
 ($L\rightarrow - V^\mu \epsilon_\mu$).

In the former expressions $\vec{\phi}$, $\Psi$, $\Psi_\Delta$, $\Psi_{N^\ast}$,
 $\Psi_{N^{\prime\ast}}$ and $A_\mu$ stand for the 
pion, nucleon, $\Delta$, $N^\ast$,
 $N^{\prime\ast}$ and photon fields, respectively 
; $N^\ast$ and $N^{\prime\ast}$ stand for the $N^\ast$(1440) and $N^\ast$(1520) 
resonances;
$m$ and $\mu$ are the nucleon
and the pion masses; $\vec{\sigma}$ and $\vec{\tau}$ 
are the spin and isospin 1/2
operators; $\vec{S^\dagger}$ and $\vec{T}^\dagger$ 
are the transition spin and isospin
operators from 1/2 to 3/2 with the normalization
\begin{equation}
\langle\frac{3}{2},M|S_\nu^\dagger|\frac{1}{2},m\rangle = C (\frac{1}{2},1,
\frac{3}{2};m,\nu,M)
\end{equation}
with $\nu$ in spherical base, and the same for $T^\dagger$. The operators 
$\vec{S}_\Delta$ and $\vec{T}_\Delta$ are the ordinary spin and isospin matrices
for the a spin and isospin 3/2 object.
For the pion fields we used the Bjorken and Drell convention :
\begin{equation}
\phi_+ = \frac{1}{\sqrt{2}}(\phi_1 - i\phi_2)\hspace{0.5cm} 
destroys \hspace{0.4cm}\pi^+, creates \hspace{0.4cm}\pi^-
\end{equation}
\begin{equation}
\phi_- = \frac{1}{\sqrt{2}}(\phi_1 + i\phi_2)\hspace{0.5cm}
 destroys \hspace{0.4cm} \pi^-, creates \hspace{0.4cm} \pi^+
\end{equation}
\begin{equation}
\phi_0 =\phi_3\hspace{0.5cm} destroys \hspace{0.4cm} \pi^0, creates 
\hspace{0.4cm} \pi^0  
\end{equation}
Hence the $|\pi^+\rangle$ state corresponds to 
$- |1 1\rangle$ in isospin base.
\\
In all formulae we have assumed that $\sigma^i \equiv\sigma_i$, $S^i\equiv
S_i$, $T^i\equiv T_i$ are Euclidean vectors
and their meaning is of a contravariant component. However 
for $\partial_i$, $A_i$, $\vec{\phi}^{(\rho)}_i$,
 $p_i$, etc, we have respected their covariant meaning.
\\

{\Large{\bf A2. Feynman Rules.}}
\\

Here we write the Feynman rules for the different vertices including
already the electromagnetic form factors, which will appear for virtual 
photons
\cite{nacher,tesina,nacher2}. In this work we consider the
electromagnetic form factors at photon point ($q^2=0$). In the Table A2 we show the
values for these form factors. We assumed the photon
with momenta $q$ as an incoming particle while the pion with momentum  
$k$ is an outgoing particle in all vertices. The momentum $p$, $p^\prime$ are
those of the baryonic states just before and after the photon absorption vertex
(or pion production vertex in eq. (55)).
\begin{center}
\begin{tabular}{|r|r|c|c|c|c|}\hline  
$F_1^N(q^2)$ &  $\rightarrow$ & $1^p$, $0^n$\\
\hline
$G_M^N(q^2)$ & $\rightarrow$ & $\mu_p$,$\mu_n$ \\
\hline
$f_\gamma(q^2)$ &  $\rightarrow$ & $f_{\Delta N\gamma}$ \\
\hline
$F_1(q^2), F_2(q^2)$ & $\rightarrow$ & $f_1, f_2$ \\
\hline
$G_1(q^2), G_2(q^2), G_3(q^2)$ & $\rightarrow$ & $g_1, g_2, g_3$\\
\hline
$G_1^\prime(q^2), G_2^\prime(q^2), G_3^\prime(q^2)$ & $\rightarrow$ &
$ g_1^\prime, g_2^\prime, g_3^\prime$\\
\hline
$F_1^\Delta, G_M^\Delta, F_c(q^2), F_A(q^2)$ & $\rightarrow$ & $1,
\mu_{\Delta}$, 1, 1 \\
\hline

\end{tabular}
\end{center}
\vspace{0.3cm}

Table A2: Value of the form factors at photon point $q^2$ =0.
\vspace{0.3cm}

\begin{equation}
V^\mu_{\gamma N N} = - ie\left\{\begin{array}{c} 
F_1^N(q^2) \\ F_1^N(q^2)[\frac{\vec{p}+\vec{p}\, {^\prime}}{2m}] + i 
\frac{\vec{\sigma}\times\vec{q}}{2m}G_M^N(q^2) \end{array} \right\}
\end{equation}
\begin{equation}
V^\mu_{\gamma N \Delta} = \sqrt{\frac{2}{3}}\frac{f_\gamma(q^2)}{m_\pi}
\frac{\sqrt{s}}{m_\Delta}
\left\{\begin{array}{c} \frac{\vec{p}_\Delta}{\sqrt{s}}
(\vec{S^\dagger}\times\vec{q}) \\
\frac{p^0_\Delta}{\sqrt{s}} [\vec{S^\dagger}\times(\vec{q}-\frac{q^0}
{p^0_\Delta}\vec{p_\Delta})] \end{array} \right\}
\end{equation}



\begin{equation}
V_{\pi N \Delta} = -\frac{f^\ast}{\mu}\vec{S^\dagger}\cdot(\vec{k}-
\frac{k^0}{\sqrt{s}}\vec{p_\Delta})T^{\lambda\dagger}
\end{equation}
\begin{equation}
V_{\pi N N} = - \frac{f}{\mu}(\vec{\sigma}\vec{k}-k^0\frac{\vec{\sigma}
(\vec{p}+\vec{p}\, ^{\prime})}{2m})\tau^\lambda
\end{equation}
\begin{equation}
V^0_{N^\ast N \gamma} = 
i\frac{\vec{q}\, ^2}{2m} F_2(q^2)
-i\vec{q}\, ^2 (1 + \frac{q^0}{2m}) F_1(q^2)
\end{equation}

\begin{eqnarray}
V^i_{N^\ast N \gamma}=
F_2(q^2)[i\vec{q}\,\frac{q^0}{2m} + (\vec{\sigma}\times\vec{q})
(1+\frac{q^0}{2m})]
\\ \nonumber & & 
\hspace{-6.5cm} 
-F_1(q^2)[i\vec{q}q^0(1+\frac{q^0}{2m})+q^2\frac{1}{2m}
(\vec{\sigma}\times\vec{q})]
\end{eqnarray}

\begin{equation}
V_{N^\ast \Delta \pi} = - \frac{g_{N^\ast \Delta \pi}}{\mu}\vec{S}^\dagger
\cdot\vec{k}T^{\lambda\dagger}
\end{equation}
\begin{equation}
V_{\Delta \Delta \pi} = - \frac{f_\Delta}{\mu}\vec{S_\Delta}\cdot\vec{k}\, 
 T_\Delta^\lambda
\end{equation}
\begin{equation}
V_{N^{\ast\prime} \Delta \pi} = -(\tilde{f}_{N^{\ast\prime} \Delta \pi}+
\frac{\tilde{g}_{N^{\ast\prime}
\Delta \pi}}{\mu^2}\vec{S^\dagger}\cdot\vec{k}\vec{S}\cdot\vec{k})
T^{\lambda\dagger}
\end{equation}
\begin{equation}
V^\mu_{\gamma \Delta \Delta} = - i\left\{\begin{array}{c}
e_\Delta F_1^\Delta(q^2) \\ e_\Delta 
F_1^\Delta(q^2)[\frac{\vec{p}+\vec{p^\prime}}
{2m_\Delta}] + i
\frac{\vec{S_\Delta}\times\vec{q}}{3m}eG_M^\Delta(q^2) \end{array} \right\}
\end{equation}
\begin{equation}
V^\mu_{\pi\pi\gamma} = -iq_\pi(k^\mu +k^{\prime\mu}) F_{\gamma\pi\pi}(q^2)
\end{equation}
\begin{equation}
V^\mu_{\Delta N \gamma \pi} = -q_\pi\frac{f^\ast}{m_\pi}T^{\lambda\dagger}
\left\{\begin{array}{c}
\, \vec{S}^\dagger\, \frac{\vec{p}_\Delta}{\sqrt{s}} \\ \vec{S}^\dagger 
\end{array} \right\} F_c(q^2)
\end{equation}

\begin{equation}
V^0_{\gamma N N^{\prime\ast}}=i(G_1(q^2) + G_2(q^2) p^{\prime 0} +
 G_3(q^2) q^0)\vec{S}^\dagger
\cdot\vec{q}
\end{equation}

\begin{eqnarray}
V^i_{\gamma N N^{\prime\ast}}=-i
[(\frac{G_1(q^2)}{2m} - G_3(q^2))(\vec{S}^\dagger\cdot\vec{q})\, \vec{q} - 
iG_1(q^2)\frac{\vec{S}^\dagger\cdot\vec{q}}{2m}(\vec{\sigma}\times\vec{q})
\\ 
\nonumber & & 
\hspace{-10cm} 
-\vec{S}^\dagger \{G_1(q^2)(q^0+\frac{\vec{q}\, ^2}{2m}) + 
G_2(q^2) p^{\prime 0} q^0 + G_3(q^2)
q^2\}]
\end{eqnarray}

\begin{equation}
V_{N^\ast N \pi} = - \frac{\tilde{f}}{\mu}\vec{\sigma}
\cdot\vec{k}\tau^\lambda
\end{equation}
\begin{equation}
V_{N^\ast N \pi\pi} = -i2\tilde{C}
\end{equation}

\begin{equation}
V^\mu_{N N \pi\gamma} = -\sqrt{2} 
q_\pi\frac{f}{\mu}\left\{\begin{array}{c} 
\frac{\vec{\sigma}(\vec{p} + \vec{p}\, ^\prime)}{2m} \\
\vec{\sigma}
\end{array} \right\}
\end{equation}
\begin{equation}
V^\mu_{N^\ast N \pi\gamma} = -\sqrt{2} 
q_\pi\frac{\tilde{f}}{\mu}
\left\{\begin{array}{c} \frac{\vec{\sigma}(\vec{p} + \vec{p}\, ^\prime)}{2m} \\
\vec{\sigma}
\end{array} \right\}
\end{equation}
\begin{equation}
V_{N^\ast p \rho^{0}[\pi^+\pi^-]}=ig_{\rho} f_{\rho} D_{\rho}
F_{\rho}(s_{\rho})\, 
\vec{S}\cdot(\vec{p}_{+}-\vec{p}_{-})
\end{equation}
\begin{equation}
V_{N^\ast p \rho^{+}[\pi^+\pi^0]}=-ig_{\rho} f_{\rho} D_{\rho}
F_{\rho}(s_{\rho})\,
\vec{S}\cdot(\vec{p}_{+}-\vec{p}_{0})\sqrt{2}
\end{equation}

\begin{equation}
V_{\gamma N \rho N}= e \frac{f_{\rho N
N}}{m_{\rho}}\sqrt{2}(\vec{\sigma}\times\vec{\epsilon}_\gamma)
\cdot\vec{\epsilon}_{\rho} 
\end{equation}
\begin{equation}
V_{\Delta^\ast\Delta \pi} = -(\tilde{f}_{\Delta^\ast \Delta \pi}+
\frac{\tilde{g}_{\Delta^\ast
\Delta \pi}}{\mu^2}\vec{S^\dagger}\cdot\vec{k}\vec{S}\cdot\vec{k})
T^{\lambda\dagger}_\Delta
\end{equation}
\begin{equation}
V^0_{\gamma N \Delta^\ast}=i(G_1^\prime(q^2) + G_2^\prime(q^2) p^{\prime 0} +
 G_3^\prime(q^2) q^0)\vec{S}^\dagger
\cdot\vec{q}
\end{equation}

\begin{eqnarray}
V^i_{\gamma N \Delta^\ast}=-i
[(\frac{G_1^\prime(q^2)}{2m} - G_3^\prime(q^2))(\vec{S}^\dagger\cdot\vec{q})\, \vec{q} - 
iG_1^\prime(q^2)\frac{\vec{S}^\dagger\cdot\vec{q}}{2m}(\vec{\sigma}\times\vec{q})
\\ 
\nonumber & & 
\hspace{-10cm} 
-\vec{S}^\dagger \{G_1^\prime(q^2)(q^0+\frac{\vec{q}\, ^2}{2m}) + 
G_2^\prime(q^2) p^{\prime 0} q^0 + G_3^\prime(q^2)
q^2\}]
\end{eqnarray}
\newpage

{\Large{\bf A3. Coupling and form factors}}
\\

%
%

{\bf Coupling constants :} 
\begin{center}
\begin{tabular}{|r|l|c|c|c|c|c|c|} \hline
\hline
$g_1$ & $g_2$ & $g_3$   \\
\hline
$0.782 m^{-1}$ & $-0.410 m^{-2}$ & $0.091 m^{-2}$   \\
\hline\hline
$g_1^\prime$ & $g_2^\prime$ & $g_3^\prime$   \\
\hline
$-0.260 m^{-1}$ & $0.270 m^{-2}$ & $(\ast)$
\footnotemark \\
\hline\hline
$f_1$ & $f_2$ & $g_{N^\ast\Delta\pi}$ \\
\hline
$-0.287 m^{-2}$ &  $ -0.067 m^{-1}$ & 2.07 \\
\hline\hline
$f$ & $f^\ast$ & $f_{\Delta N\gamma}$   \\
\hline
1 & 2.13 & 0.12 \\
\hline\hline
$f_{\Delta}$ & $e$ & $f_{\rho}$ \\
\hline
0.802 & 0.3027 & 6.14 \\
\hline\hline
$\tilde{f}_{N^{\ast\prime}\Delta\pi}$ & $\tilde{g}_{N^{\ast\prime}\Delta\pi}$
& $g_{\rho}$ \\
\hline
$-1.061$ & 0.640 & $5.09$ \\
\hline\hline
$\tilde{f}_{\Delta^\ast\Delta\pi}$ & $\tilde{g}_{\Delta^\ast\Delta\pi}$
& $g_{\rho}^\prime$ \\
\hline
$-1.325$ & 0.146 & $2.60$ \\
\hline\hline
$C_{\rho}$ & $\tilde{C}$ & $\tilde{f}$ \\
\hline
3.96 & $-2.29\mu^{-1}$ & 0.477\\
\hline\hline



\end{tabular}
\end{center}
\footnotetext{We do not consider this constant which is related with
 the $S_{1/2}$ scalar
helicity amplitude, because it is not contributing 
in the photoproduction reaction.}


{\bf Form factors :}
\\*[0.5cm]

For the off-shell pions we use a form factor of the monopole type :
\begin{equation}
F_\pi(p^2) = \frac{\Lambda_\pi^2 - \mu^2}{\Lambda_\pi^2-p^2}\hspace{1.5cm}
;\, \Lambda_\pi=1250\, MeV
\end{equation}

The used for factor for the off-shell $\rho$ meson is :
\begin{equation}
F_{\rho}(q^2) = \frac{\Lambda_{\rho}\, ^2 - m_{\rho}^2}{\Lambda_{\rho}\, ^2 -q^2}
\end{equation}
with $\Lambda_{\rho}$ = 1.4 GeV. We have tested our results using different form
factors and we have found small changes in the results. 
\\

{\Large{\bf A4. Amplitudes for the reaction}}
\\

In this appendix we write the explicit expressions for the amplitudes of the
Feynman diagrams used in the model. The isospin coefficients and some constant
factors are collected in the coefficients $C$ which are written in the table A4.
In the following expressions $q$, $p_1$, $p_2$, $p_4$, and $p_5$ are the momentum of the 
photon,the incoming nucleon, the outgoing nucleon and the two pions :
\\
\\*[0.5cm]
\begin{center}
\begin{tabular}{|r|r|c|c|c|c|}\hline  
$\gamma$ & $p$  & $\rightarrow$ & $\pi^+$ & $\pi^-$ & $p$\\
\hline
$q$ & $p_1$ &  & $p_5$ & $p_4$ & $p_2$ \\
\hline
\end{tabular}
\\*[0.4cm]
\begin{tabular}{|r|r|c|c|c|c|}\hline
$\gamma$ & $p$  & $\rightarrow$ & $\pi^+$ & $\pi^0$ &
$n$\\
\hline
$q$ & $p_1$ &  & $p_5$ & $p_4$ & $p_2$ \\
\hline
\end{tabular}
\\*[0.4cm]
\begin{tabular}{|r|r|c|c|c|c|}\hline
$\gamma$ & $p$  & $\rightarrow$ & $\pi^0$ & $\pi^0$ & $p$\\
\hline
$q$ & $p_1$ &  & $p_5$ & $p_4$ & $p_2$ \\
\hline
\end{tabular}
\end{center}

\begin{tabular}{|r|l|c|c|c|c|c|c|} \hline
D. & $\pi^+\pi^-p$ & $\pi^+\pi^0 n$ & $\pi^0\pi^0 p$ & D. 
& $\pi^+\pi^-p$ & $\pi^+\pi^0 n$ & $\pi^0\pi^0 p$\\
\hline\hline
$\bf{(a)}$ & $2i$ & -$i\sqrt{2}$ & 0 & $\bf{(l)}$ & $i/3$ & $i\sqrt{2}/3$ & $i2/3$ \\
\hline
$\bf{(a^\prime)}$ & 0 & 0 & 0 & $\bf{(l^{\, \prime})}$ & $i$ & -$i\sqrt{2}/3$ & 
$i2/3$ \\
\hline\hline
$\bf{(b)}$ & -$2i$ & 0 & 0 & $\bf{(m)}$ & $i/9$ & $i \sqrt{2}/9$ & $i 2/9$ 
 \\
\hline
$\bf{(b^\prime)}$ & 0 & $i\sqrt{2}$ & 0 & 
$\bf{(m^\prime)} $& $i/3$ & -$i \sqrt{2} /9$ 
&  $i 2/9$\\
\hline\hline
$\bf{(c)}$ & 2$i$ & -$i \sqrt{2}$ & 0 & $\bf{(o)}$ & -$2/3$ & -2$ i \sqrt{2}$ /3 & -1/3 
 \\
\hline
$\bf{(c^\prime)}$ & 0 & 0  & 0 & $\bf{(o^\prime)}$ & 1 & $\sqrt{2}$/3 & -1/3\\
\hline\hline
$\bf{(d)}$ & -2$i$ & 0 & 0 & $\bf{(p)}$ & 0 & 0 & $i$2/9 \\
\hline
$\bf{(d^\prime)}$ & 0 & $i\sqrt{2}$  & 0 & $\bf{(p^\prime)}$ & $i/3$ & -$i\sqrt{2}/9$ & $i 2/9$  \\
\hline\hline
$\bf{(e)}$ & 2$i$ & -$i\sqrt{2}$ & $i$ & $\bf {(q)}$ & -1/9 & -$\sqrt{2}$/9 & -2/9  \\
\hline
$\bf{(e^\prime)}$ & 0 & $i \sqrt{2}$  & $i$ & $\bf{(q^\prime)}$ & -1/3 & $\sqrt{2}$/9 & -2/9  \\
\hline\hline
$\bf{(f)}$ & 2$i$ & -$i \sqrt{2}$ & $i$ & $\bf{(r)}$ & 2 & 2 & 2    \\
\hline
$\bf{(f^\prime})$ & 0 & $i \sqrt{2}$  & $i$ & & & &  \\
\hline\hline
$\bf{(g)}$ & 2$i$ & -$i\sqrt{2}$ & $i$ & $\bf{(s)}$ & 2$i$ & -$i \sqrt{2} $
 & 0 \\
\hline
$\bf{(g^\prime)}$ & 0 & $i\sqrt{2}$  & $i$ &  $\bf{(s^\prime)}$ & 0 & 0 & $0$\\ 
\hline\hline
$\bf{(h)}$ & $i$2/9 & -$i\sqrt{2}$/9 & -$i$2/9 & $\bf{(t)}$ & -2$i$ & 0 & $0$ \\
\hline
$\bf{(h^\prime)}$ & 0 & -2$i\sqrt{2}$/9 & -$i$2/9 &  $\bf{(t^\prime)}$ & 0 & 
$i\sqrt{2}$ &
$0$ \\
\hline\hline
$\bf(i)$ & $i/9$ & $i\sqrt{2}/9$ & $0$   
& $\bf{(u)}$ & $i\sqrt{2/3}$ & $i2/\sqrt{3}$ & $i\sqrt{1/6}$ \\
\hline
$\bf{(i^\prime)}$ & -$i/3$ & 0   & $0$  
& $\bf{(u^\prime)}$ & $-i\sqrt{3/2}$ & $-i/(2\sqrt{3})$ & $i\sqrt{1/6}$\\
\hline\hline
$\bf{(j)}$ & $i/9$ & $i\sqrt{2}$/9 & 0 & $\bf{(v)}$ & $-i$ & $i\sqrt{2}$ & 0 \\
\hline
$\bf{(j^\prime})$ & -$i/3$ & 0   & 0 & $\bf{(w)}$ & 0 & 1 & 0 \\
\hline\hline
$\bf{(k)}$ & -$i$2/9 & -2$i \sqrt{2}$/9 & -$i$2/9  & $\bf{(x)}$ & $-i\sqrt{2/3}$ &
$i\sqrt{1/3}$ & 0 \\
\hline
$\bf{(k^\prime)}$ & 0 & $i \sqrt{2}$/9 & -$i$2/9  & & & & \\

\hline
\end{tabular}




\vspace{1cm}

\vspace{0.3cm}

Table A4: Coefficients of the amplitudes for the $\gamma p\rightarrow\pi\pi N$
 reactions, accounting
for isospin and constant factors.
\vspace{0.3cm}


We write only the amplitude when the pion labelled $p_5$ is
emitted before the pion labelled $p_4$, except
in the cases where only one possibility is avalaible 
and the explicit amplitudes for these cases is written. We have also
evaluated the crossed diagrams when the pion labelled $p_5$ is emitted 
after the pion called $p_4$.
Such amplitudes are exactly the same than the others written before, but
exchanging the momenta $p_4$ and $p_5$ and changing some isospin coefficient.
 This latter change is taken into account by the factor $C$ called
 with a label $\prime$ written in table
 A4.

We should note that in the vertex  $\Delta N \pi$, when $\vec{p}_\Delta$ is
not zero, we must change $\vec{p}_\pi$ by $\vec{p}_\pi - \frac{p^0}
{\sqrt{s}}\vec{p}_\Delta$ for the final pion.

In the formulae, $D_\pi$, $D_\rho$, $G_\Delta$, $G_N$, $G_{N^\ast}$, $G_{N^{\prime\ast}}$
$G_{\Delta^\ast}$ 
are the propagator of the pion, rho, delta, nucleon, $N^\ast(1440)$,  
$N^\ast(1520)$, $\Delta(1700)$ respectively. Expressions for them and for the width of the 
resonances can be found in \cite{tejedor,tejedor2,cano} and 
in the present work for some of them.
The labels in the amplitudes and coefficents in the Table A4 are making
reference to the diagrams in the fig. 1 except in the amplitudes and coefficents
called $u,v,w,x$. These ones belongs to the diagrams showed in the figs. $7b,
2a, 2b, 7a$ respectively.

\begin{eqnarray}
-i T_a^\mu & = & Ce(\frac{f}{\mu})^2 G_N(p_2 + p_4)\frac{\Lambda^2 
-\mu^2}{\Lambda^2 - (p_5 - q)^2} \\
& & \times [-p_4^0\frac{\vec{\sigma}\cdot(2\vec{p}_2+\vec{p}_4)}{2m} + 
\vec{\sigma}\cdot\vec{p_4}] F_A(q^2)\nonumber   \\
& & \times\left\{\begin{array}{c} 
 \frac{\vec{\sigma}(2\vec{p_1} + \vec{q} - \vec{p_5})}{2m} \\ \nonumber
\vec{\sigma} \nonumber
\end{array} \right\}
\end{eqnarray}
\begin{eqnarray}
-i T_b^\mu & = & Ce(\frac{f}{\mu})^2 G_N(p_1 - p_5)\frac{\Lambda^2 
-\mu^2}{\Lambda^2 - (p_4 - q)^2} \\
& & \times\left\{\begin{array}{c} 
 \frac{\vec{\sigma}\cdot(2\vec{p_2} - \vec{q} + \vec{p_4})}{2m} \\ \nonumber
\vec{\sigma} \nonumber 
\end{array} \right\} \\ \nonumber
& & \times [-p_5^0\frac{\vec{\sigma}(2\vec{p}_1-\vec{p}_5)}{2m} + 
\vec{\sigma}\cdot\vec{p_5}] F_A(q^2)
\end{eqnarray}

\begin{eqnarray}
-i T_c^\mu & = & Ce(\frac{f}{\mu})^2 G_N(p_2 + p_4)D_\pi(p_5 - q)
F_{\gamma\pi\pi}(q^2)  \\
& & \times \frac{\Lambda^2 -\mu^2}{\Lambda^2 - (p_5 - q)^2}
[-p_4^0\frac{\vec{\sigma}(2\vec{p}_2+\vec{p}_4)}{2m} + 
\vec{\sigma}\cdot\vec{p_4}]\nonumber   \\
& & \times [-(p_5 - q)^0\frac{\vec{\sigma}\cdot
(\vec{p}_1+\vec{p}_2+\vec{p}_4)}{2m} + 
\vec{\sigma}\cdot(\vec{p_5} - \vec{q})]\nonumber 
 \\  & & \times\left\{\begin{array}{c}  
2p_5 - q \nonumber
\end{array} \right\}^\mu 
\end{eqnarray}
\begin{eqnarray}
-i T_d^\mu & = & Ce(\frac{f}{\mu})^2 G_N(p_1 - p_5)D_\pi(p_4 - q)
F_{\gamma\pi\pi}(q^2)  \\
& & \times \frac{\Lambda^2 -\mu^2}{\Lambda^2 - (p_4 - q)^2}
[-(p_4 - q)^0\frac{\vec{\sigma}\cdot(\vec{p}_1 - \vec{p}_5 + \vec{p}_2)}{2m} + 
\vec{\sigma}\cdot(\vec{p_4}-\vec{q})]\nonumber   \\
& & \times [-p_5^0\frac{\vec{\sigma}\cdot
(2\vec{p}_1-\vec{p}_5)}{2m} +
\vec{\sigma}\cdot\vec{p_5}]\nonumber
\\  & & \times\left\{\begin{array}{c}  
2p_4 - q \nonumber
\end{array} \right\}^\mu 
\end{eqnarray}
\begin{eqnarray}
-i T_e^\mu & = & Ce(\frac{f}{\mu})^2 G_N(p_2 + p_4)G_N(p_1 + q)  \\
& & \times [-p_4^0\frac{\vec{\sigma}\cdot(2\vec{p}_2+\vec{p}_4)}{2m} + 
\vec{\sigma}\cdot\vec{p_4}]\nonumber   \\
& & \times [-p_5^0\frac{\vec{\sigma}\cdot
(\vec{p}_2+\vec{p}_4)}{2m} + 
\vec{\sigma}\cdot\vec{p_5}]\nonumber 
 \\  & & \times\left\{\begin{array}{c}  
F_1^p(q^2) \\ \nonumber
F_1^p(q^2)[\frac{\vec{p}+\vec{p} ^{\, \prime}}{2m}] + 
iG_M^p(q^2)\frac{\vec{\sigma}
\times\vec{q}}{2m} \nonumber
\end{array} \right\} \\ \nonumber
\end{eqnarray}
\begin{eqnarray}
-i T_f^\mu & = & Ce(\frac{f}{\mu})^2 G_N(p_2 + p_4)G_N(p_1 - p_5)  \\
& & \times [-p_4^0\frac{\vec{\sigma}\cdot(2\vec{p}_2+\vec{p}_4)}{2m} + 
\vec{\sigma}\vec{p_4}]\nonumber \\
& & \times\left\{\begin{array}{c}
F_1^N(q^2) \\ \nonumber
F_1^N(q^2)[\frac{\vec{p}+\vec{p} ^{\, \prime}}{2m}] +
iG_M^N(q^2)\frac{\vec{\sigma}
\times\vec{q}}{2m} \nonumber
\end{array} \right\} \\ \nonumber
& & \times [-p_5^0\frac{\vec{\sigma}\cdot
(2\vec{p}_1-\vec{p}_5)}{2m} +
\vec{\sigma}\vec{p_5}]\nonumber
\end{eqnarray}
\begin{eqnarray}
-i T_g^\mu & = & Ce(\frac{f}{\mu})^2 G_N(p_2 - q)G_N(p_1 - p_5)  \\
& & \times\left\{\begin{array}{c}
F_1^N(q^2) \\ \nonumber
F_1^N(q^2)[\frac{\vec{p}+\vec{p} ^{\, \prime}}{2m}] +
iG_M^N(q^2)\frac{\vec{\sigma}
\times\vec{q}}{2m} \nonumber
\end{array} \right\} \\ \nonumber
& & \times [-p_4^0\frac{\vec{\sigma}\cdot
(\vec{p}_1-\vec{p}_5+\vec{p}_2-\vec{q})}{2m} +
\vec{\sigma}\cdot\vec{p_4}]\nonumber \\
& & \times [-p_5^0\frac{\vec{\sigma}\cdot
(2\vec{p}_1-\vec{p}_5)}{2m} +
\vec{\sigma}\cdot\vec{p_5}]\nonumber
\end{eqnarray}
\begin{equation}
-iT_h^0 = 0 \hspace{0.4cm} in \hspace{0.3cm}\gamma-p\hspace{0.3cm}CM
\hspace{0.3cm}frame
\end{equation}
\begin{eqnarray}
-i T_h^i & = & C\frac{f}{\mu}\frac{f^\ast}{\mu}
\frac{f_\gamma(q^2)}{\mu}
G_N(p_2 + p_4)G_\Delta(p_1 + q) \\ 
& & \times [-p_4^0\frac{\vec{\sigma}\cdot(2\vec{p}_2+\vec{p}_4)}{2m} + 
\vec{\sigma}\cdot\vec{p_4}]\nonumber  \\
& & \times
[-2i(\vec{p}_5\times\vec{q})
-(\vec{\sigma}\cdot\vec{q})\vec{p}_5
+(\vec{p}_5\cdot
\vec{q})\cdot\vec{\sigma}]\frac{p_\Delta^0}{m_\Delta}
\end{eqnarray}
\begin{eqnarray}
-i T_i^\mu & = & Ce(\frac{f^\ast}{\mu})^2
G_\Delta(p_2 + p_4) F_\pi((p_5-q)^2)
F_c(q^2)  \\ 
& & \times\left\{\begin{array}{c}
[2\vec{p}_4\cdot\vec{p}_5 - i(\vec{p}_4\times\vec{p}_5)\cdot\vec{\sigma}]
\frac{-1}{\sqrt{s_\Delta}}  \\ \nonumber
2\vec{p}_4 - i (\vec{\sigma}\times\vec{p}_4)
\end{array} \right\}
\end{eqnarray}
\begin{eqnarray}
-i T_j^\mu & = & Ce(\frac{f^\ast}{\mu})^2
G_\Delta(p_2 + p_4)D_\pi(p_5 - q) F_\pi((p_5-q)^2)
F_{\gamma\pi\pi}(q^2) \\ \nonumber
& & \times 
[2\vec{p}_4\cdot(\vec{p}_5-\vec{q}) - i(\vec{p}_4\times(\vec{p}_5-\vec{q}))
\cdot\vec{\sigma}] \\ \nonumber
& & \times\left\{\begin{array}{c}  
2p_5 - q \nonumber
\end{array} \right\}^\mu 
\end{eqnarray}
\begin{eqnarray}
-i T_k^\mu & = & C\frac{f}{\mu}\frac{f^\ast}{\mu}
\frac{f_\gamma(q^2)}{\mu}
G_N(p_2 + p_4)G_\Delta(p_1 + q) \\ 
& & \times\left\{\begin{array}{c}
[-2i(\vec{p}_4\times\vec{q})
-(\vec{\sigma}\cdot\vec{q})\vec{p}_4
+(\vec{p}_4\cdot
\vec{q})\vec{\sigma}]\frac{\vec{p}_\Delta}{m_\Delta}
 \\ \nonumber
[-2i(\vec{p}_4\times\vec{q}^{\, \prime} )
-(\vec{\sigma}\cdot\vec{q}^{\, \prime} )\vec{p}_4
+(\vec{p}_4\cdot
\vec{q}^{\, \prime} )\vec{\sigma}]\frac{p_\Delta^0}{m_\Delta}
\end{array} \right\} \\ \nonumber
& & \times [-p_5^0\frac{\vec{\sigma}(2\vec{p}_1-\vec{p}_5)}{2m} + 
\vec{\sigma}\vec{p_5}]
\end{eqnarray}
with $\vec{q}^{\, \prime} =
( \vec{q} - \frac{q^0}{p_\Delta^0}\vec{p_\Delta} )$
\\

{\bf Amplitude of $N^{\prime\ast}$(1520):}
\\
Vector part :
\\
\begin{eqnarray}
-iT_l^i & = &
C\frac{f^\ast}{\mu}G_\Delta(p_2 + p_4)G_{N^{\prime\ast}}(p_1 + q) \\
\nonumber
& &
\times\vec{S}\cdot\vec{p}_4 [\tilde{f}_{N^{\prime\ast}\Delta\pi}
+ \frac{\tilde{g}_{N^{\prime\ast}\Delta\pi}}{\mu^2}
\vec{S}^\dagger\cdot\vec{p}_5\vec{S}\cdot\vec{p}_5] \\ 
\nonumber
 & & \times\{(\frac{G_1(q^2)}{2m} - G_3(q^2))(\vec{S}^\dagger\cdot\vec{q})\, 
 \vec{q} - 
i G_1(q^2)\frac{\vec{S}^\dagger\cdot\vec{q}}{2m}(\vec{\sigma}\times\vec{q}) \\
\nonumber
& & 
-\vec{S}^\dagger [G_1(q^2)(q^0+\frac{\vec{q}\, ^2}{2m}) + 
G_2(q^2) p^{\prime 0} q^0 + G_3(q^2)
q^2]\}
\end{eqnarray}
\\
Scalar part:

\begin{eqnarray}
-iT_l^0 & = &
-C\frac{f^\ast}{\mu}G_\Delta(p_2 + p_4)G_{N^{\prime\ast}}(p_1 + q) \\
\nonumber
& &
\times\vec{S}\cdot\vec{p}_4 [\tilde{f}_{N^{\prime\ast}\Delta\pi}
+ \frac{\tilde{g}_{N^{\prime\ast}\Delta\pi}}{\mu^2}
\vec{S}^\dagger\cdot\vec{p}_5\vec{S}\cdot\vec{p}_5] \\ 
\nonumber
& &
\times [G_1(q^2) + G_2(q^2) p^{\prime 0} 
+ G_3(q^2) q^0] \vec{S}^\dagger\cdot\vec{q}
\nonumber
\end{eqnarray}

\begin{eqnarray}
-i T_m^\mu & = & C e(\frac{f^\ast}{\mu})^2 G_N(p_1 + k)G_\Delta(p_2 +
p_4) F_\pi((p_5-q)^2) \\ \nonumber
& & \times[2\vec{p}_4\cdot\vec{p}_5 - 
i(\vec{p}_4\times\vec{p}_5)\cdot\vec{\sigma}] \\ \nonumber
& & \times\left\{\begin{array}{c}
F_1^p(q^2) \\ \nonumber
F_1^p(q^2)[\frac{\vec{p}+\vec{p} ^{\, \prime}}{2m}] + iG_M^p(q^2)\frac{\vec{\sigma}
\times\vec{q}}{2m} \nonumber
\end{array} \right\} \\ \nonumber
\end{eqnarray}

\begin{equation}
-iT_o^0 = 0 \hspace{0.4cm} in \hspace{0.3cm}\gamma-p\hspace{0.3cm}CM
\hspace{0.3cm}frame
\end{equation}
\begin{eqnarray}
-i T_o^i & = & C \frac{f^\ast}{\mu}\frac{f_\Delta}{\mu}\frac{f_\gamma(q^2)}
{\mu}G_\Delta(p_2 + p_4)G_\Delta(p_1 + q) \\ \nonumber
& & \times [i\frac{5}{6}(\vec{p}_4\cdot\vec{q})\vec{p}_5 - 
i\frac{5}{6}(\vec{p}_5\cdot\vec{q})\vec{p}_4 -
\frac{1}{6}(\vec{p}_4\cdot\vec{p}_5)(\vec{\sigma}\times\vec{q})-\\ \nonumber
& & \frac{1}{6}(\vec{p}_4\cdot\vec{\sigma})(\vec{p}_5\times\vec{q}) +
\frac{2}{3}(\vec{p}_5\cdot\vec{\sigma})(\vec{p}_4\times\vec{q})] \nonumber
\end{eqnarray}
\begin{eqnarray}
-i T_p^0 & = & C(\frac{f^\ast}{\mu})^2G_\Delta(p_2 + p_5)G_\Delta(p_1 - p_4)
F_\pi((p_5-q)^2)
\{e_\Delta F_1^\Delta(q^2) \\ \nonumber
& & 
\times[2\vec{p}_5\cdot\vec{p}_4 
- i(\vec{p}_5\times\vec{p}_4)\cdot\vec{\sigma}]\} \\ \nonumber
\end{eqnarray}

\begin{eqnarray}
-i T_p^i & = & C(\frac{f^\ast}{\mu})^2G_\Delta(p_2 + p_5)G_\Delta(p_1 - p_4)
F_\pi((p_5-q)^2) \\ 
\nonumber
& & 
\times\{\frac{e_\Delta F_1^\Delta(q^2)}{2}
\frac{(\vec{p}_1-2\vec{p}_4)}{m_\Delta}
[2\vec{p}_5\cdot\vec{p}_4 
- i(\vec{p}_5\times\vec{p}_4)\cdot\vec{\sigma}]+ \\ \nonumber
& & i\frac{eG_M^\Delta(q^2)}{m}[i\frac{5}{6}(\vec{p}_4\cdot\vec{q})\vec{p}_5 -
i\frac{5}{6}(\vec{p}_5\cdot\vec{q})\vec{p}_4 -
\frac{1}{6}(\vec{p}_5\times\vec{q})(\vec{p}_4\cdot\vec{\sigma})-\\ \nonumber
& & \frac{1}{6}(\vec{p}_5\cdot\vec{\sigma})(\vec{p}_4\times\vec{q}) +
\frac{2}{3}(\vec{p}_5\cdot\vec{p}_4)(\vec{\sigma}\times\vec{q})]\} \nonumber
\end{eqnarray}

{\bf Amplitude of $N^{\ast}$(1440):}
\\
Vector part :
\\
\begin{eqnarray}
-iT_q^i & = & C 
\frac{f^\ast}{\mu}\frac{g_{N^{\ast}\Delta\pi}}{\mu}
G_\Delta(p_2 + p_4) G_{N^\ast}(p_1 + q)\vec{S}\cdot
\vec{p}_4\vec{S}^\dagger\cdot\vec{p}_5 \\ 
& & \nonumber\times\
\{F_2(q^2)[i\vec{q}\,\frac{q^0}{2m}+(\vec{\sigma}\times\vec{q})
(1+\frac{q^0}{2m})]
\\ \nonumber & & 
-F_1(q^2)[i\vec{q}q^0(1+\frac{q^0}{2m})+q^2\frac{1}{2m}
(\vec{\sigma}\times\vec{q})]\}
\end{eqnarray}
\\
Scalar part:

\begin{eqnarray}
-iT_q^0 & = & C 
\frac{f^\ast}{\mu}\frac{g_{N^{\ast}\Delta\pi}}{\mu}\
G_\Delta(p_2 + p_4) G_{N^\ast}(p_1 + q) \vec{S}\cdot
\vec{p}_4\vec{S}^\dagger\cdot\vec{p}_5\\ 
& & \nonumber \times\
\{i\frac{\vec{q}\, ^2}{2m} F_2(q^2)
-i\vec{q}\, ^2 (1 + \frac{q^0}{2m}) F_1(q^2)\}
\end{eqnarray}

\begin{eqnarray}
-iT_r^\mu & = & C\tilde{C}\frac{{\tilde{f}_\gamma(q^2)}}{\mu}G_{N^\ast}(p_1 + q)
 \left\{\begin{array}{c} 
0 \\ 
(\vec{\sigma}\times\vec{q})
\end{array} \right\}
\end{eqnarray}
\begin{eqnarray}
-i T_s^\mu & = & Ce(\frac{\tilde{f}}{\mu})^2 G_{N^\ast}(p_2 + p_4)\frac{\Lambda^2 
-\mu^2}{\Lambda^2 - (p_5 - q)^2}F_A(q^2) \\
& & \times (\vec{\sigma}\cdot\vec{p_4})\left\{\begin{array}{c} 
 \frac{\vec{\sigma}\cdot(2\vec{p_1} + \vec{q} - \vec{p_5})}{2m} \\ \nonumber
\vec{\sigma} \nonumber
\end{array} \right\}
\end{eqnarray}
\begin{eqnarray}
-i T_t^\mu & = & Ce(\frac{\tilde{f}}{\mu})^2 G_{N^\ast}(p_1 - p_5)\frac{\Lambda^2
-\mu^2}{\Lambda^2 - (p_4 - q)^2}F_A(q^2) \\
& & \times (\vec{\sigma}\cdot\vec{p_5})\left\{\begin{array}{c}
 \frac{\vec{\sigma}\cdot(2\vec{p_2} - \vec{q} + \vec{p_4})}{2m} \\ \nonumber
\vec{\sigma} \nonumber
\end{array} \right\}
\end{eqnarray}
{\bf Amplitude of $\Delta(1700)$:}
\\
Vector part :
\\
\begin{eqnarray}
-iT_u^i & = &
C\frac{f^\ast}{\mu}G_\Delta(p_2 + p_4)G_{\Delta^\ast}(p_1 + q) \\
\nonumber
& &
\times\vec{S}\cdot\vec{p}_4 [\tilde{f}_{\Delta^\ast\Delta\pi}
+ \frac{\tilde{g}_{\Delta^\ast\Delta\pi}}{\mu^2}
\vec{S}^\dagger\cdot\vec{p}_5\vec{S}\cdot\vec{p}_5] \\ 
\nonumber
 & & \times\{(\frac{G_1^\prime(q^2)}{2m} - G_3^\prime(q^2))(\vec{S}^\dagger\cdot\vec{q})\, 
 \vec{q} - 
i G_1^\prime(q^2)\frac{\vec{S}^\dagger\cdot\vec{q}}{2m}(\vec{\sigma}\times\vec{q}) \\
\nonumber
& & 
-\vec{S}^\dagger [G_1^\prime(q^2)(q^0+\frac{\vec{q}\, ^2}{2m}) + 
G_2^\prime(q^2) p^{\prime 0} q^0 + G_3^\prime(q^2)
q^2]\}
\end{eqnarray}
\\
Scalar part:

\begin{eqnarray}
-iT_u^0 & = &
-C\frac{f^\ast}{\mu}G_\Delta(p_2 + p_4)G_{\Delta^\ast}(p_1 + q) \\
\nonumber
& &
\times\vec{S}\cdot\vec{p}_4 [\tilde{f}_{\Delta^\ast\Delta\pi}
+ \frac{\tilde{g}_{\Delta^\ast\Delta\pi}}{\mu^2}
\vec{S}^\dagger\cdot\vec{p}_5\vec{S}\cdot\vec{p}_5] \\ 
\nonumber
& &
\times [G_1^\prime(q^2) + G_2^\prime(q^2) p^{\prime 0} 
+ G_3^\prime(q^2) q^0] \vec{S}^\dagger\cdot\vec{q}
\nonumber
\end{eqnarray}  
{\bf Amplitude for $N^\ast(1520)\rightarrow\rho N$}
\\
Vector part :
\\
\begin{eqnarray}
-iT_v^i & = &
-C g_{\rho} f_{\rho}D_{\rho}(p_4+p_5) F_{\rho}(s_{\rho})
G_{N^{\ast\prime}}(p_1 + q)
\vec{S}\cdot(\vec{p}_{+}-\vec{p}_{-}) \\
\nonumber
 & & \times\{(\frac{G_1(q^2)}{2m} - G_3(q^2))(\vec{S}^\dagger\cdot\vec{q})\, 
 \vec{q} - 
i G_1(q^2)\frac{\vec{S}^\dagger\cdot\vec{q}}{2m}(\vec{\sigma}\times\vec{q}) \\
\nonumber
& & 
-\vec{S}^\dagger [G_1(q^2)(q^0+\frac{\vec{q}\, ^2}{2m}) + 
G_2(q^2) p^{\prime 0} q^0 + G_3(q^2)
q^2]\}
\end{eqnarray}
\\
Scalar part:

\begin{eqnarray}
-iT_v^0 & = &
C g_{\rho} f_{\rho}D_{\rho}(p_4+p_5) F_{\rho}(s_{\rho})
G_{N^{\ast^\prime}}(p_1 + q) 
\vec{S}\cdot(\vec{p}_{+}-\vec{p}_{-})\\
\nonumber
& &
\times [G_1(q^2) + G_2(q^2) p^{\prime 0} 
+ G_3(q^2) q^0] \vec{S}^\dagger\cdot\vec{q}
\nonumber
\end{eqnarray}  
\begin{equation}
-iT_{w}=-C e \sqrt{2} f_{\rho} 
\frac{f_{\rho N N}}{m_{\rho}} D_{\rho} F_{\rho}(s_{\rho})
(\vec{\sigma}
\times\vec{\epsilon}_\gamma)\cdot(\vec{p}_+
-\vec{p}_0)
\end{equation}

{\bf Amplitude for $\Delta(1700)\rightarrow\rho N$}
\\
Vector part :
\\
\begin{eqnarray}
-iT_x^i & = &
-C g_{\rho}^\prime f_{\rho}D_{\rho}(p_4+p_5) F_{\rho}(s_{\rho})
G_{\Delta^\ast}(p_1 + q)
\vec{S}\cdot(\vec{p}_{+}-\vec{p}_{-}) \\
\nonumber
 & & \times\{(\frac{G_1^\prime(q^2)}{2m} - G_3^\prime(q^2))(\vec{S}^\dagger\cdot\vec{q})\, 
 \vec{q} - 
i G_1^\prime(q^2)\frac{\vec{S}^\dagger\cdot\vec{q}}{2m}(\vec{\sigma}\times\vec{q}) \\
\nonumber
& & 
-\vec{S}^\dagger [G_1^\prime(q^2)(q^0+\frac{\vec{q}\, ^2}{2m}) + 
G_2^\prime(q^2) p^{\prime 0} q^0 + G_3^\prime(q^2)
q^2]\}
\end{eqnarray}
\\
Scalar part:

\begin{eqnarray}
-iT_x^0 & = &
C g_{\rho}^\prime f_{\rho}D_{\rho}(p_4+p_5) F_{\rho}(s_{\rho})
G_{\Delta^\ast}(p_1 + q) 
\vec{S}\cdot(\vec{p}_{+}-\vec{p}_{-})\\
\nonumber
& &
\times [G_1^\prime(q^2) + G_2^\prime(q^2) p^{\prime 0} 
+ G_3^\prime(q^2) q^0] \vec{S}^\dagger\cdot\vec{q}
\nonumber
\end{eqnarray}

\end{document}